\documentclass{article}
\usepackage{graphicx}
\usepackage[numbers]{natbib}
\usepackage{xcolor}
\usepackage{amsbsy, amssymb, amsthm, amsmath, dsfont}
\usepackage{comment}
\usepackage{indentfirst}
\usepackage[utf8]{inputenc}

\usepackage{float}
\usepackage{url}

\newtheorem{theorem}{Theorem}
\newtheorem{proposition}[theorem]{Proposition}

\usepackage{tikz}
\usetikzlibrary{shapes.geometric}
\usetikzlibrary{trees}

\usepackage[linesnumbered,ruled,vlined]{algorithm2e}

\usepackage{environ}
\newcounter{parentAlgoLine}
\SetKwBlock{BEGIN}{}{}
\makeatletter
\NewEnviron{substeps}{%
  \protected@edef\theparentequation{\theequation}%
  \setcounter{parentAlgoLine}{\value{AlgoLine}}%
  \setcounter{AlgoLine}{0}%
  \BEGIN{\BODY}%
  \setcounter{AlgoLine}{\value{parentAlgoLine}}%
  \ignorespacesafterend
}
\makeatother

\makeatletter
\DeclareFontEncoding{LS1}{}{}
\DeclareFontSubstitution{LS1}{stix}{m}{n}
\DeclareMathAlphabet{\mathscr}{LS1}{stixscr}{m}{n}
\makeatother

\makeatletter
\DeclareFontEncoding{LS1}{}{}
\DeclareFontSubstitution{LS1}{stix}{m}{n}
\DeclareMathAlphabet{\mathscrr}{LS1}{stixscr}{m}{n}

\usepackage{setspace}
\title{Nonparametric inference for nonstationary spatial point processes}

\author{
Izabel Nolau\thanks{Departamento de Métodos Estatísticos, Instituto de Matemática e Estatística, Universidade Federal do Rio de Janeiro, Rio de Janeiro, Brazil, \texttt{nolau@dme.ufrj.br}},
Flávio B. Gonçalves\thanks{Departmento de Estatística, Universidade Federal de Minas Gerais, Belo Horizonte, Brazil, \texttt{fbgoncalves@ufmg.br}} \hspace{0.001cm} and Dani Gamerman\thanks{Departamento de Métodos Estatísticos, Instituto de Matemática e Estatística, Universidade Federal do Rio de Janeiro, Rio de Janeiro, Brazil, \texttt{dani@im.ufrj.br}}}
\date{}

\begin{document}

\maketitle

\begin{abstract}
Point pattern data often exhibit features such as abrupt changes, hotspots and spatially varying dependence in local intensity. Under a Poisson process framework, these correspond to discontinuities and nonstationarity in the underlying intensity function. These features are difficult to capture with standard modeling approaches. This paper proposes a spatial Cox process model in which nonstationarity is induced through a random partition of the spatial domain, with conditionally independent Gaussian process priors specified across the resulting regions. This construction allows for heterogeneous spatial behavior, including sharp transitions in intensity. A discretization-free MCMC algorithm is developed to target the infinite-dimensional posterior distribution without approximation, thus ensuring exact inference. The random partition framework via Voronoi tessellation also reduces the computational burden associated with Gaussian process models. Spatial covariates can be incorporated to account for structured variation in intensity. The proposed methodology is evaluated through synthetic examples and real-world applications, demonstrating its ability to flexibly capture complex spatial structures. The model performs competitively, outperforming stationary and nonstationary alternatives in a variety of scenarios. Recent computational methods are used, enabling scalability to large datasets while preserving exactness. The paper concludes with a discussion of potential extensions and directions for future work.
\end{abstract}

\noindent%

\section{Introduction}

Spatial point processes are stochastic models used to explain the variation of point pattern data, where the locations of occurrences are observed over a region of interest \citep{diggle2014statistical}. Unlike geostatistical or areal data, where the sample space consists of vectors of fixed or bounded dimension, each realization of a point pattern is a set of locations of random cardinality, with each element lying in a continuous region of $\mathbb{R}^d$. The Poisson process is one of the most widely used models in this context, with its intensity function (IF, hereafter) describing the instantaneous rate of occurrence across the spatial domain \citep{moller2004statistical}. Cox processes \citep{cox1955some} let the IF vary stochastically, accommodating a wide range of dependence structures; among these, nonparametric formulations in which the IF is a continuous function of a latent Gaussian process (GP) are particularly flexible and have become a standard tool for capturing complex, irregular, or application-specific spatial behavior \citep{moller1998log, moller2007modern, illian2012toolbox, gonccalves2018exact, gonccalves2023corrigendum}.

\begin{figure}[!ht]
    \centering
    \includegraphics[width=\linewidth]{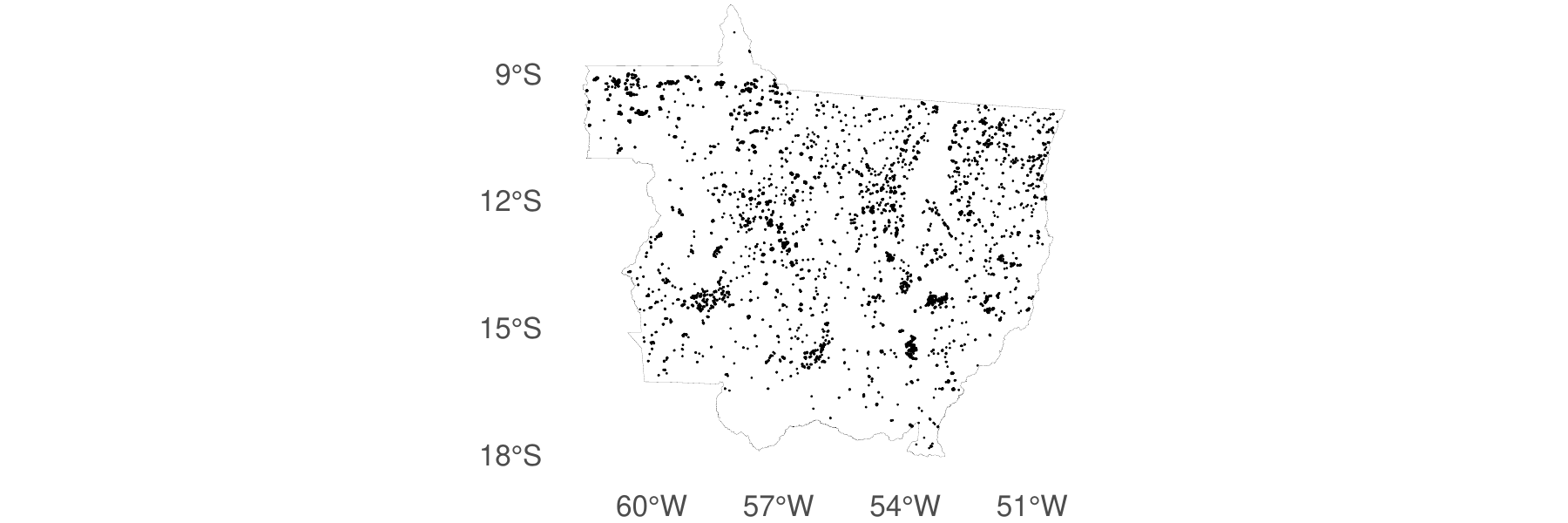}
    \caption{Observed fire occurrences during the dry season of 2023 (May to August) in Mato Grosso, Brazil.}
    \label{fig__motivation}
\end{figure}

Although point process modeling initially focused on stationary and homogeneous formulations \citep{ripley1977modelling, diggle1985kernel, baddeley2000non}, stationarity is rarely tenable in practice. Figure \ref{fig__motivation} illustrates fire occurrences in the state of Mato Grosso, Brazil, a large and environmentally diverse region where highly concentrated hotspots coexist with vast areas of sparse or nearly absent activity. Patterns of this kind are governed by underlying drivers, such as biomes, land use, and administrative boundaries, that change abruptly in space, producing intensity surfaces with markedly different local regimes and sharp transitions between them. Stationary models cannot represent such behavior, and even standard nonstationary models tend to oversmooth, blurring the very discontinuities that characterize the phenomenon.

Capturing this behavior requires a nonstationary covariance structure, and herein lies the central difficulty. Estimating flexible nonstationary covariance functions is hard even in the geostatistical setting, where the GP is directly observed and replications may be available \citep{sampson1992nonparametric, paciorek2006spatial}. In a GP-driven Cox process the task is considerably more demanding: the GP is latent, it enters the model nonlinearly through the IF of a Poisson process, and there is a single realization of the process, with no replication to inform the spatial structure. Compounding this, the likelihood of a GP-driven Cox process is intractable, as it requires integrating an infinite-dimensional GP against a Poisson likelihood that admits no closed form. The overwhelming majority of the literature circumvents this intractability through approximation, most commonly by discretizing the domain into a grid and replacing the continuous IF by a piecewise-constant surrogate, but also through Laplace approximations or numerical quadrature \citep{moller1998log, brix2001spatiotemporal, diggle2013spatial, liang2009analysis, pinto2015point, illian2012toolbox}. Such approximations introduce an error that is difficult to quantify and whose propagation through the posterior is poorly understood (see \cite{lee2023unified, nguyen2024independent} for a discussion in the context of Bayesian nonparametrics); controlling it demands increasingly fine grids, with costs that scale cubically in the number of cells, and can distort or erase structural features of the IF \citep{silva2024exact, hildeman2018level}. These shortcomings are especially damaging under nonstationarity, where the features of interest are precisely the local regimes and sharp transitions that discretization tends to smear.

The few works that pursue nonstationary GP-driven Cox processes inherit, and in some respects aggravate, these limitations. To keep the covariance estimable from a single point pattern, \cite{dvorak2019quick} introduce nonstationarity through smooth, covariate-driven parametric specifications of the mean and covariance of the log-Gaussian field, and estimate it with a three-step composite-likelihood procedure based on first- and second-order summary statistics. \cite{dangelo2022local} adopt a local log-Gaussian Cox process in which the GP parameters vary smoothly in space, fitted by local Palm likelihood over moving windows. Both deliver smoothly varying structures that cannot represent the abrupt changes seen in Figure \ref{fig__motivation}, and both rely on multi-step, ad-hoc estimation that departs from a full likelihood-based treatment and yields only frequentist, summary-statistic-based uncertainty rather than coherent uncertainty quantification for the IF and its features. In short, existing approaches sacrifice either flexibility, exactness, or robust uncertainty quantification, and typically more than one.

A parsimonious yet flexible route to nonstationarity is to model the process as locally stationary over a partition of the domain, with conditionally independent stationary GPs across regions \citep{kim2005analyzing, luo2024nonstationary}. Treating the partition as random accounts for partition uncertainty and, crucially, induces conditional independence that confines inference to subdomains rather than the full domain, with substantial computational savings; it also accommodates sharp changes in the IF across region boundaries while retaining the well-understood machinery of stationary GPs within regions.

Voronoi tessellations are especially attractive in this regard, combining flexibility with parsimony \citep{heikkinen1998non, pope2021gaussian}, alongside product/tree-based partitions \citep{luo2024nonstationary, Lu06082025} and level-set constructions \citep{dunlop2017hierarchical, hildeman2018level, gonccalves2023exactlevelset}. Most of this body of work has been developed for geostatistical data, where the field is discretely observed. The partition-based constructions that do address point processes remain limited in precisely the directions that matter here: those of \cite{heikkinen1998non}, \cite{gonccalves2023exactlevelset}, and \cite{Lu06082025} restrict the IF to be piecewise constant, while \cite{hildeman2018level} relies on discrete approximations of both the partition-defining field and the GPs driving the IF. A flexible, continuously varying GP-driven nonstationary IF with exact inference has therefore remained an open problem.

This paper covers that gap. It proposes a Cox process whose IF is a continuous function of conditionally independent stationary GPs defined over the regions of a random Voronoi partition of the spatial domain, with the partition treated as unknown and its uncertainty propagated to the posterior. The formulation accommodates heterogeneous local behavior, including abrupt changes and discontinuities across regions, while both stationary and nonstationary patterns are handled within a single unified model. Inference is carried out with a Markov chain Monte Carlo algorithm that targets the exact posterior of all unknown components, including the local intensities, the within-region correlation structure, and the partition itself. Unlike the prevailing literature, the method involves no discretization and no grid refinement, and is therefore free of approximation error; the partition structure further localizes the cubic cost of GP computations to individual subdomains. Because inference is exact and fully Bayesian, it delivers coherent and robust uncertainty quantification for every structural feature of the model. Finally, we exploit a valid sparse GP construction \citep{goncalves2025pcnngp} to scale the methodology to large datasets while preserving exactness. To the best of our knowledge, this is the first GP-driven Cox process to combine flexible nonstationary intensity estimation, the ability to capture sharp spatial changes, and exact, discretization-error-free Bayesian inference with robust uncertainty quantification, offering a substantial methodological advance for point process analysis under nonstationarity.

The remainder of the paper is organized as follows. Sections 2 and 3 describe the proposed model and inference methodology, respectively. Section 4 presents numerical experiments with both simulated and real datasets. Section 5 concludes with a discussion of the results and potential directions for future research.

\section{Model Specification} \label{sec__model_specification}

Let $\mathcal{Y}$ be a Poisson process in $\mathcal{S}$ with IF $\lambda(s): \mathcal{S} \to \mathbb{R}^+$, where $\mathcal{S}$ is some compact region in $\mathbb{R}^d$. The sample space of $\mathcal{Y}$ consists of all finite sets of points in $\mathcal{S}$.  Let $S = (S_1, \cdots, S_L)$ be an unknown partition of $\mathcal{S}$ defined by a tessellation of convex polytopes. The following Cox process model is assumed for $\mathcal{Y}$:
\begin{eqnarray}
\mathcal{Y} \mid \lambda &\sim& PP_{\mathcal{S}}(\lambda), \label{eq__model__1}\\
\lambda(s) &=& \sum_{\ell=1}^L \lambda^*_\ell F(\beta_\ell(s))\mathds{1}(s \in S_\ell), \qquad \mbox{for } \ell=1, \cdots, L ,\label{eq__model__2}
\end{eqnarray}
where $PP_{\mathcal{S}}(\lambda)$ denotes a Poisson process on $\mathcal{S}$ with IF $\lambda$, for all $\mathcal{S} \subset \mathbb{R}^d$. Equation \eqref{eq__model__2} implies that the IF in region $S_\ell$ is $\lambda^*_\ell F(\beta_\ell(s))$. This formulation allows the global intensity $\lambda(s)$ to capture heterogeneous behavior and discontinuities or edges between regions. It is assumed that $F: \mathcal{S} \to [0, 1]$ is a known monotone function, so that $\lambda^*_\ell$ is an upper bound of the IF in $S_\ell$.
The $\beta_\ell$ components are Gaussian processes in $\mathcal{S}$ that induce local smooth variation of the intensity within each region. The same structure adopted for each region $S_\ell$ in \eqref{eq__model__2} is used in other works, but for the whole region $\mathcal{S}$. For example, in \cite{gonccalves2018exact, gonccalves2023corrigendum} with $F$ being the standard Gaussian c.d.f. and \cite{adams2009tractable} with $F$ being the sigmoid function.

Under model in (\ref{eq__model__1})-(\ref{eq__model__2}), and conditional on  $\lambda$, the $Y_\ell = \mathcal{Y} \bigr|_{S_\ell}$ ($\mathcal{Y}$ restricted to $S_\ell$) are independent Poisson processes in $S_\ell$ with intensity $\lambda^*_\ell F(\beta_\ell(s))$ \citep[see][section 5.1]{kingman1992poisson}. Therefore, a valid likelihood for $\lambda$ \citep[see][Section 4.3]{gonccalves2024likelihood}, based on an observation $\mathscrr{y}$ of $\mathcal{Y}$, is given by
\begin{eqnarray}\label{eq__likelihood}
\pi(\mathscrr{y} \mid \lambda^*, \beta, S) &=& \prod_{\ell = 1}^L \left[\exp\left\{- \int_{S_\ell} \lambda^*_\ell F(\beta_\ell(s)) ds\right\} \prod_{s \in y_\ell} \lambda^*_\ell F(\beta_\ell(s))\right],
\end{eqnarray}
where $\lambda^* = (\lambda^*_1, \cdots, \lambda^*_L)$, $\beta = (\beta_1, \cdots, \beta_L)$ and $y_\ell=\mathscrr{y}\bigr|_{S_\ell}$. The $\ell$-th product term on the r.h.s. of \eqref{eq__likelihood} is the unnormalized density of $Y_\ell$ w.r.t. the measure of a $PP_{\mathcal{S}}(1)$, known as the Poisson process likelihood. Here, the process $Y_\ell$ is actually defined as the Poisson process with intensity $\lambda$ in $S_\ell$ and zero in $\mathcal{S}\setminus S_\ell$ so that the dominating measure is independent of the partition.

The likelihood function in \eqref{eq__likelihood} is intractable because the integral of $F(\beta_{\ell})$ is not analytically available and this is often handled through approximations in the literature. For instance, discretizing $\mathcal{S}$ introduces bias and potentially leads to serious model mischaracterization. To address this issue, a novel model augmentation formulation that yields a tractable augmented likelihood function is considered, enabling exact inference.

\subsection{The Augmented Model}\label{sec__augmented_model}

Our augmented model extends the Poisson thinning scheme in \cite{adams2009tractable} and \cite{gonccalves2018exact, gonccalves2023corrigendum} to account for the random partition structure of the proposed model. 

Three Poisson processes associated with each region $S_\ell$ are defined: 
\begin{eqnarray} \label{eq__augmented_model__start}
Y_\ell \mid \lambda^*_\ell, \beta_\ell, S_\ell
&\sim& PP_{S_\ell}\left(\lambda^*_\ell F(\beta_\ell(\cdot))\right),\\
\tilde{Y}_\ell \mid \lambda^*_\ell, \beta_\ell, S_\ell
&\sim& PP_{S_\ell}\left(\lambda^*_\ell [1 - F(\beta_\ell(\cdot))]\right),
\label{eq__augmented_model__tilde} \\
 Z_\ell \mid \lambda^*_\ell, S_\ell &\sim& PP_{\mathcal{S} \setminus S_\ell}(\lambda^*_\ell), \label{eq__augmented_model__end}
\end{eqnarray}
where all the processes $Y_\ell$, $\tilde{Y}_\ell$, and $Z_\ell$ are conditionally independent, given the IF.

\begin{figure}[ht!]
    \centering
    \includegraphics[height = 4.5cm]{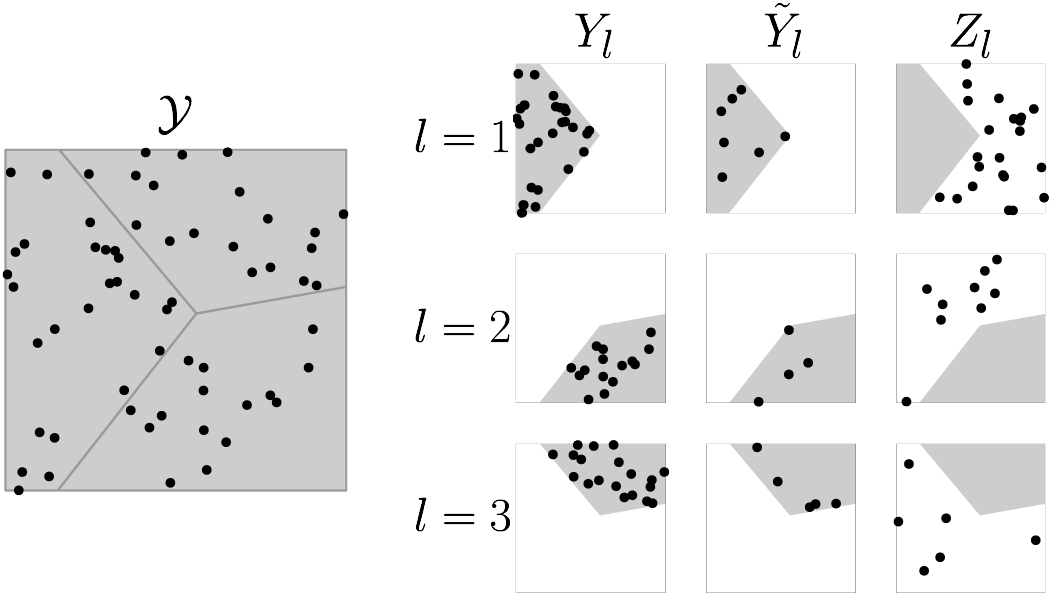}
    \caption{Illustration of data $\mathcal{Y}$ (left panel) and latent processes $(Y, \tilde Y, Z)$ for $L = 3$ (right panel). The shaded areas represent each region $S_\ell$.}
    \label{fig__augment_3__simulation}
\end{figure}

The Colouring Theorem \citep[see][section 5.1]{kingman1992poisson} implies that the superposition $(Y_\ell \cup \tilde{Y}_\ell \cup Z_\ell)$ is a homogeneous Poisson process on $\mathcal{S}$ with IF $\lambda^*_\ell$. Furthermore, the first two processes can be viewed as the retained and thinned events, respectively, resulting from a Poisson thinning of $PP_{S_{\ell}}(\lambda^*_\ell)$ with probabilities $F(\beta_{\ell})$.
Figure \ref{fig__augment_3__simulation} illustrates the augmented model for $L=3$.

The observed data satisfies $\mathcal{Y} = \bigcup_{\ell = 1}^L Y_\ell$, and each process $Y_\ell$, being a deterministic function of $\mathcal{Y}$ and $S_1,\ldots,S_L$, is unknown since the sets $S_1,\ldots,S_L$ are unknown.

Define $Y = (Y_1, \cdots, Y_L)$, $\tilde{Y} = (\tilde{Y}_1, \cdots, \tilde{Y}_L)$ and $Z = (Z_1, \cdots, Z_L)$ and let $y_\ell$, $\tilde{y}_\ell$, and $z_\ell$ be a realization of the respective processes. 
The augmented likelihood function is given by the joint density of $(\mathcal{Y} , Y, \tilde Y, Z)$. Given the conditional independence of the processes and the Poisson likelihood formula, standard calculations lead to tractable form
\begin{eqnarray} \label{eq__augmented_likelihood}
\pi(\mathscrr{y}, y, \tilde{y}, z \mid \cdot) \propto \mathds{1}\left(\mathscrr{y} = \bigcup_{\ell = 1}^L y_\ell \right)\prod_{\ell = 1}^L \left[\lambda_\ell^{*T_\ell} \exp\left\{- \lambda^*_\ell |\mathcal{S}| \right\} \prod_{s \in y_\ell} F(\beta_\ell(s))\prod_{s \in \tilde{y}_\ell} [1 - F(\beta_\ell(s))]\right],
\end{eqnarray}
where $T_\ell$ is the cardinality of $y_\ell \cup \tilde{y}_\ell \cup z_\ell$, and $|\mathcal{S}|$ is the volume of $\mathcal{S}$.

The augmented model preserves the original model in \eqref{eq__model__1}-\eqref{eq__model__2} upon marginalization over the augmented variables and the induced augmented likelihood function in \eqref{eq__augmented_likelihood} allows the derivation of an exact inference methodology.
The derivation leading to Equation \eqref{eq__augmented_likelihood} is given in online Appendix A.1.

\subsection{Prior Distribution}\label{sec__prior_distribution}

Let $\theta_\ell$ denote the hyperparameter vector indexing the prior of $\beta_\ell$, and define $\theta = (\theta_1, \cdots, \theta_L)$. It is assumed, \emph{a priori}, that each $\beta_\ell$ is an independent GP defined on the spatial domain $\mathcal{S}$, and the spatial partition is determined by a Voronoi tessellation induced by a set of generator points $U = (U_1, \cdots, U_L) \subset \mathbb{R}^d$ \citep{aurenhammer2013voronoi}.

Gaussian processes are a flexible and widely used prior for modeling unknown functions, including IFs of Poisson processes. In our setting, assigning distinct hyperparameters $\theta_\ell$ to each region enhances model flexibility, enabling features such as region-specific smoothness. Each GP $\beta_\ell$ is assumed to be stationary with mean $\mu_\ell$, variance $\sigma_\ell^2$, and isotropic correlation function $\rho(h)$, where $h = \|s - s'\|$ denotes the Euclidean distance between locations $s, s' \in \mathcal{S}$. There are many possible options for the correlation function. This work adopts the power exponential family defined as
\begin{eqnarray}\label{eq__correlation_function}
\rho(h) \ = \ \exp\left\{ - \frac{h^{\gamma}}{2\phi} \right\}, \qquad \text{with } \phi > 0 \text{ and } 0 < \gamma \leq 2.
\end{eqnarray}

Details about inference for $\theta$ are provided in Section \ref{sec__updating_beta}. Prior sensitivity analyses reported in \cite{gonccalves2018exact, gonccalves2023corrigendum} suggests that inference is robust if $\theta$ is fixed at reasonable values, deduced from the scale of the spatial domain $\mathcal{S}$ and the properties of the link function $F$. For example, when $F$ is the standard Gaussian c.d.f. $\Phi$, or a rescaled sigmoid function (which closely approximates $\Phi$), the choice $\mu_\ell = 0, \sigma_\ell^2 = 4$ is typically appropriate \citep[see][]{gonccalves2018exact, gonccalves2023corrigendum}. Parameter $\gamma$, which controls the smoothness of the sample paths, is difficult to estimate even under discretely observed GPs, and is therefore fixed at a plausible value (commonly between 1.5 and 2). The range parameter $\phi$ also governs smoothness and should ideally be estimated from the data. However, in the latent GP structure of the model, it is also difficult to estimate. This issue can be attenuated by fixing the value of $\phi$ or by using a strongly informative prior, such as a truncated Uniform or a Gamma distribution, based on the spatial scale of $\mathcal{S}$ and the desired resolution of the intensity estimation (provided sufficient data information is available). In practice, $\phi$ is often chosen relative to the maximum spatial extent of $\mathcal{S}$ (e.g., the diagonal of its bounding hypercube), which provides a natural reference to control the smoothness of the process: larger $\phi$ yields smoother intensities, while smaller $\phi$ allows for more erratic behavior.

Voronoi tessellations offer a simple yet flexible way to partition space. Given a set of generator points $\{U_1, \ldots, U_L\} \subset \mathbb{R}^d$, the Voronoi tessellation divides $\mathcal{S}$ into $L$ subregions, where each cell contains the locations closest to its corresponding generator. The partition is thus fully characterized by $U$, which is estimated alongside the other model unknowns. Because the partition plays a central role in our approach, the prior on $U$ must be chosen with care. In preliminary experiments, we observed label switching and identifiability issues stemming from the nonuniqueness of the mapping between $U$ and the resulting partition -- difficulties well known in models with group structures and exchangeable components \citep[see][]{fruhwirth2011dealing}. We therefore adopt a repulsive joint prior that penalizes proximity between generator points. Following \cite{quinlan2021class}, we adopt the repulsive prior:
\begin{eqnarray}\label{eq__U_prior}
\pi(u) & \propto & \frac{1}{|\mathcal{S}|^L} \, \prod_{1 \leq \ell_1 \leq \ell_2 \leq L} \left(1 - \exp\left\{-\eta \, || \, u_{\ell_1} - u_{\ell_2} \, || ^\nu\right\}\right), \qquad 
\end{eqnarray}
where $\eta > 0$ and $\nu > 0$ are fixed hyperparameters that control the strength and shape of the repulsion, respectively. Practical guidance on choosing $\eta$ and $\nu$ is provided in the online Appendix D.1.

Since $F$ maps to $(0,1)$, the parameter $\lambda^*_\ell$ is an upper
bound of the intensity function (IF) in region $S_\ell$. Identifying
this parameter is a delicate issue due to its multiplicative interaction
with the link function $F$. Ideally, $\lambda^*_\ell$ should be
identified as the approximate maximum of the IF over $S_\ell$. This is
facilitated by the model and prior specifications: the monotonicity and
scale of $F$ (for example $F=\Phi$), combined with the mean $0$ and variance $4$ of the GP
$\beta_\ell$, encourage $\beta_\ell$ to attain values near $2$, which
in turn identifies $\lambda^*_\ell$ as the approximate maximum of the
IF in $S_\ell$. In more extreme situations, such as regions with few
observations or a nearly constant IF, the prior on $\lambda^*_\ell$
can provide additional regularization by penalizing large values of the
parameter. A Gamma prior is particularly appealing here, as it is conjugate to the augmented model likelihood.


The partition structure serves two key purposes: it allows the model to capture spatial heterogeneity and improves computational efficiency by enabling conditional localized inference. Since the number of regions $L$ does not necessarily correspond to a physical quantity, it may be treated as a user-specified tuning parameter. Moreover, the blocking scheme is designed to ensure a valid and efficient algorithm and would be compromised by transdimensional approaches, such as reversible-jump MCMC, required to estimate $L$. Choosing $L$ involves a trade-off between model flexibility and parsimony. Practical guidance on choosing $L$ is provided in the online Appendix D.2. Empirical insights are provided in Section \ref{sec__applications}. Finally, although the primary goal is spatial adaptation and scalability, the induced partitions can also serve a secondary purpose as a clustering mechanism for spatial point data.

\subsection{Model Properties}

In spatial models, the dependence structure plays a crucial role in characterizing the relationships among locations. The following properties formally explain the model's ability to capture spatial heterogeneity. In order to derive them, this is redefined in terms of a single process $\beta_0$ as
\begin{equation*}
\lambda ( s ) \ = \ \lambda_0^* (s) F ( \beta_0 (s) ),  \quad \mbox{with} \quad 
\lambda_0^* (s) \ = \ \sum_{\ell=1}^L \lambda^*_\ell \mathds{1}(s \in S_\ell) \quad \mbox{and} \quad
\beta_0 ( s ) \ = \ \sum_{\ell=1}^L  \beta_\ell(s)
\mathds{1}(s \in S_\ell).
\end{equation*}

The marginal and crossed moments for the latent process $\beta_0$ and the IF $\lambda$ are obtained by integrating out the random partition through $U$. Defining $\kappa_\ell(s) = P(s \in S_\ell)$, for $\ell = 1, \cdots, L$, it follows that, for all $s\in\mathcal{S}$,
\begin{eqnarray*}
&\displaystyle E( \beta_0 (s) \mid \theta ) \ = \ \sum_{\ell = 1}^L \kappa_\ell(s)\mu_\ell,\quad 
Var(\beta_0 (s) \mid \theta) \ = \ \sum_{\ell = 1}^L\kappa_\ell(s)\sigma^2_\ell, & \\
&\displaystyle E( \lambda (s) \mid \theta ) \ = \ \sum_{\ell = 1}^L \kappa_\ell(s) \lambda_{\ell}^*\ E [ F ( \beta_\ell (s)) \mid U , \theta ].&
\end{eqnarray*}

{\proposition \label{proposition__cov_beta}
For $s, s' \in \mathcal{S}$, define $\kappa_\ell(s,s')=P(s \in S_\ell, s' \in S_\ell)$, for $\ell = 1, \cdots, L$, and $\kappa_{\ell_1, \ell_2}(s,s')=P(s \in S_{\ell_1}, s' \in S_{\ell_2})-P(s \in S_{\ell_1})P(s' \in S_{\ell_2})$, for $\ell_1, \ell_2 = 1, \cdots, L$, then
\begin{eqnarray}
\nonumber Cov(\beta_0 (s), \beta_0 (s') \mid \theta)
&=& \sum_{\ell = 1}^L \kappa_\ell(s,s')\sigma^2_\ell\rho(h) \ + \
\sum_{\ell = 1}^L\sum_{m = 1}^L \kappa_{\ell_1, \ell_2}(s,s')\mu_\ell\mu_m. \label{equation__cov_beta}
\end{eqnarray}
If $\beta_0$ is a zero mean process, that is, $\mu_\ell = 0$, for $\ell = 1, \cdots, L$, then
\begin{eqnarray}\label{eq__cov_beta}
Cov(\beta_0 (s), \beta_0 (s') \mid \theta) &=&
\sum_{\ell = 1}^L \kappa_\ell(s,s')\sigma^2_\ell\rho(h).
\end{eqnarray}
}

{\proposition \label{proposition__cov_lambda}
For $s, s' \in \mathcal{S}$,
\begin{eqnarray}
&& Cov(\lambda(s), \lambda(s') \mid \lambda^*, \theta) \nonumber\\
&=&E\left[\sum_{\ell = 1}^L\lambda^{*2}_\ell \ Cov\left(F(\beta_\ell(s)), F(\beta_\ell(s')) \ \middle| \ \lambda^*, U\right)\mathds{1}(s \in S_\ell)\mathds{1}(s' \in S_\ell) \ \middle| \ \lambda^*, \theta\right] \ + \nonumber \\
&&Cov\left(\sum_{\ell_1 = 1}^L\lambda^*_{\ell_1} E\left[F(\beta_{\ell_1}(s)) \ \middle| \ \lambda^*, U\right]\mathds{1}(s \in S_{\ell_1}), \right.
\nonumber \\
&& \qquad \ \, \left. \sum_{\ell_2 = 1}^L\lambda^*_{\ell_2}E\left[F(\beta_{\ell_2}(s')) \ \middle| \ \lambda^*, U\right]\mathds{1}(s' \in S_{\ell_2}) \ \middle| \ \lambda^*, \theta\right)
\label{eq__cov_lambda}
\end{eqnarray}
}

Finally, 
\begin{eqnarray*}
Var(\lambda(s) \mid \lambda^*, \theta) &=& \sum_{\ell = 1}^L\lambda^{*2}_\ell \ E\left[Var\left(F(\lambda_\ell(s)) \ \middle| \ \lambda^*, U\right)\mathds{1}(s \in S_\ell) \ \middle| \ \lambda^*, \theta \right] \ + \\
&& \sum_{\ell = 1}^L\lambda^{*2}_\ell \ Var\left(E\left[F(\lambda_\ell(s)) \ \middle| \ \lambda^*, U\right]\mathds{1}(s \in S_\ell) \ \middle| \ \lambda^*, \theta \right).
\end{eqnarray*}

The proofs of Propositions \ref{proposition__cov_beta} and \ref{proposition__cov_lambda} are provided in the online Appendix A.2 and A.3. The above results and the fact that the prior on $U$ results in nonstationary functions $\kappa_\ell(s,s')$, imply a nonstationary covariance function for the IF. This feature will be illustrated in Section
\ref{sec__applications}.

\subsection{Incorporating Spatial Covariates}

Spatial covariates can be incorporated into the IF by replacing $\beta_\ell(s)$ with $\beta_\ell(s) + W(s)'\alpha_\ell$ in (\ref{eq__model__2}), where $W(s) = (W_1(s), \cdots, W_p(s))'$ is a vector of $p$ spatially indexed covariates observed at $s$, $\forall s \in \mathcal{S}$, and $\alpha_\ell = (\alpha_{1, \ell}, \cdots, \alpha_{p, \ell})'$ is the corresponding vector of regression coefficients, which are assumed to be constant within each region. A simpler submodel is obtained by setting $\alpha_\ell = \alpha, \forall \ell$. Both formulations provide interpretable ways to incorporate covariates while retaining the spatial structure of the predictor.

A typical choice for the prior of $\alpha_\ell$ is a zero mean multivariate normal distribution.

\section{Bayesian Inference}\label{sec__bayesian_inference}

This section addresses the estimation of the IF of the proposed model from a single realization $\mathscrr{y}$ of the Cox process. The vector of unknown model components is given by
\begin{equation}\label{eq__parametric_vector}
\psi \ = \ \left(Y, \tilde{Y}, Z, \lambda^*, \beta, \theta, U\right),
\end{equation}
where each element corresponds to latent variables or parameters introduced in the model specification.

Inference is carried out within the Bayesian framework, targeting the posterior distribution $\pi(\psi \mid \mathscrr{y})$. The augmented model presented in Section \ref{sec__augmented_model} facilitates this task by yielding an analytically tractable augmented likelihood. The posterior measure $P$ is absolutely continuous with respect to the prior measure $Q$, and is given by Bayes’ theorem as
\[
\frac{dP}{dQ}(\psi) \ \propto \ \pi(\mathscrr{y} \mid \psi).
\]

All components in $\psi$ admit prior densities with respect to suitable dominating measures, with the exception of the latent Gaussian process $\beta$, whose dominating measure has to depend on $\theta$, for example, the prior GP measure itself. Therefore, conditional on $\theta$, it follows that
\begin{eqnarray}\label{eq__posterior}
&\pi(\psi \mid \mathscrr{y}, \theta) \ = \
\pi(\mathscrr{y} \mid y) \,
\pi(y \mid \lambda^*, \beta, u) \,
\pi(\tilde{y} \mid \lambda^*, \beta, u) \,
\pi(z \mid \lambda^*, u) \, 
\pi(\beta \mid \theta) \,
\pi(\lambda^*) \,
\pi(u),&
\end{eqnarray}
where the product of the first four densities on the r.h.s. is given in expression \eqref{eq__augmented_likelihood}.
As a result, the unnormalized full conditional densities of all components in $\psi$, except $\theta$, are proportional to \eqref{eq__posterior}.

Given the complexity of the posterior, particularly its infinite-dimensional nature, inference is performed via Markov Chain Monte Carlo (MCMC). The proposed algorithm targets the exact posterior distribution of $\psi$, ensuring that the only source of approximation is the usual MCMC error.

\subsection{MCMC Algorithm} \label{section__comp}

A Gibbs sampling algorithm is developed for posterior inference under the model. Its blocking scheme exploits the conditional independence of the model components given the partition. The sampler consists of five blocks: one updates the random partition, and the remaining four each update a set of components within the partition regions independently, namely, the latent point processes, the GPs $\beta_\ell$, the parameters $\lambda^*_\ell$, and the GP hyperparameters.

Each block is sampled from its full conditional distribution either directly or via a Metropolis-Hastings (MH) step. The partition update is designed to yield a valid algorithm without resorting to transdimensional methods such as reversible jump MCMC. The following blocks are defined:
\begin{equation}\label{eq__blocks}
(\tilde{Y}, Z), \quad (U, Y, \tilde{Y}, Z),  \quad \beta, \quad \lambda^*, \quad (\theta,\beta_{\mathcal{S} \setminus y^*}).
\end{equation}

The latent processes \(\tilde{Y}\) and \(Z\) appear in two blocks, but their union is held fixed in the second block while only their labels are updated. This is necessary to avoid inconsistencies with the partition \(U\), and fixing the dimension of their union is crucial to the validity of this step. The component $\beta_{\mathcal{S} \setminus y^*}$ represents the infinite-dimensional remainder of $\beta$, introduced later in the algorithm and included in the final MCMC block to enable valid updates of the hyperparameter $\theta$. This component is sampled retrospectively, that is, its values are only unveiled (i.e. simulated) at a finite, albeit random, set of locations required to carry out the MCMC steps. This retrospective mechanism is essential to ensure that the algorithm targets the exact posterior distribution of $\psi$, despite the infinite-dimensional nature of the latent process. Other instances where retrospective sampling is employed to avoid finite-dimensional approximations can be found in \cite{beskos2006exact}, \cite{papaspiliopoulos2008retrospective} and \cite{gonccalves2023exactmontecarlo}.

Nonhomogeneous Poisson processes with bounded IFs are efficiently sampled via Poisson thinning \citep{lewis1979simulation}, presented in Algorithm 1.

\begin{algorithm}[H]
\caption{Poisson thinning algorithm to simulate a \(PP_A(\lambda)\), with \(\lambda(s)\leq\lambda^*, \forall s\in A.\)}\label{alg__simulate_ipp}

Draw \(K \sim \text{Poisson}(\lambda^* |A|)\);

Draw \(v_k\sim U(A)\) independently for all \(k = 1, \cdots, K\);

Retain each \(v_k\) independently with probability \(\lambda(v_k)/\lambda^*\);

Return the retained locations.
\end{algorithm}

The algorithms for updating each block in the Gibbs sampler are summarized below.

\subsubsection{Updating $(\tilde{Y}, Z)$} 

This step updates the latent locations associated with the processes $(\tilde{Y}_\ell, Z_\ell)$, for $\ell = 1, \cdots, L$, introduced to achieve likelihood tractability under the augmented model formulation. Given \(\mathscrr{y}\) and the remaining components of \(\psi\), the variables \(\tilde{Y}_\ell\) and \(Z_\ell\) are conditionally independent with respective distributions given by Equations \eqref{eq__augmented_model__tilde} and \eqref{eq__augmented_model__end}. Sampling is performed via Algorithm \ref{alg__simulate_ipp}. To sample $\tilde{Y}_\ell$, the Gaussian process $\beta_\ell$ must be sampled retrospectively at the proposed locations, since the retention probabilities in Step 3 depend on its values at those points.

\subsubsection{Updating $(U, Y, \tilde{Y}, Z)$} 

This block jointly updates the partition $U$ and the processes $(Y, \tilde{Y}, Z)$, as their configuration depends on the induced partition of $\mathcal{S}$. This updated is a MH step. The union \(\left(\bigcup_{\ell=1}^L \tilde{Y}_\ell\right) \cup \left(\bigcup_{\ell=1}^L Z_\ell\right)\) is held fixed, and only the labels are updated. The component \(Y\) is updated to maintain consistency with the partition \(U\), since \(\bigcup_{\ell=1}^L Y_\ell = \mathcal{Y}\).

A proposal \((u, y, \tilde{y}, z)\rightarrow(\ddot{u}, \ddot{y}, \ddot{\tilde{y}}, \ddot{z})\) is drawn from:
\begin{equation}\label{eq__proposal__varphi}
q(\ddot{u}, \ddot{y}, \ddot{\tilde{y}}, \ddot{z} \mid u, y, \tilde{y}, z) \ = \ q(\ddot{\tilde{y}}, \ddot{z} \mid \ddot{u}, \tilde{y}, z) \, q(\ddot{y}\mid \ddot{u}, y) \, q(\ddot{u} \mid u).
\end{equation}

For some fixed \(b\leq L\), the algorithm updates \(U_{\mathscrr{\ell}^*}\) and its \(b\) closest neighbors, where \(\mathscrr{\ell}^*\) is uniformly drawn from \(\{1, \cdots, L\}\). This localized update improves chain mixing. Proposals are generated via a mixture of two uniform distributions on balls centered at current values, with the smaller-radius component having higher weight (e.g., \(> 0.9\)). The purpose of the larger-radius component is to promote occasional long jumps, reducing the chances of the chain becoming stuck in local modes of the posterior distribution.

Given \(\ddot{U}\), the observations \(\mathscrr{y}\) are consistently relabeled to define \(\ddot{Y}\). For points in \(\tilde{Y} \cup Z\) that change region between \(U\) and \(\ddot{U}\), the label reassignment is performed. A location $s$ that changes from $S_k$ to $S_\ell$ is allocated to $\tilde{Y}_\ell$ with probability:
\begin{equation}\label{eq__allocation_probability}
p_{\ell k}(s) \ = \ \frac{\lambda^*_\ell [1 - F(\beta_\ell(s))]}{\lambda^*_k + \lambda^*_\ell [1 - F(\beta_\ell(s))]},   
\end{equation}
and is allocated to $Z_k$ with probability $1-p_{\ell k}(s)$. This proposal is designed to preserve chain reversibility and maximize the acceptance probability by being consistent with the prior distribution of $(\tilde{Y},Z)$.

The acceptance probability is:
\begin{equation}\label{eq__acceptance_probability__varphi}
1 \ \wedge \ \prod_{\ell = 1}^L \left[
\left(\lambda_\ell^*\right)^{\ddot{T}_\ell - T_\ell}
\prod_{s\in y_\ell |_{\mathcal{S} \setminus \ddot{S}_\ell}} \frac{F(\beta_{\ddot{\ell}}(s))}{F(\beta_\ell(s))}
\prod_{\substack{k = 1 \\ k \neq \ell}}^L \frac{\prod_{s \in \ddot{c}_{\ell k}} [1 - F(\beta_{\ddot{\ell}}(s))] \; \ddot{q}_{\ell k}(s)}{\prod_{s \in c_{\ell k}} [1 - F(\beta_\ell(s))] \; q_{\ell k}(s)}
\right] \frac{\pi(\ddot{u})}{\pi(u)},
\end{equation}
where \(q_{\ell k}(s) = p_{\ell k}(s)^{\mathds{1}(s \in \ddot{\tilde{y}}_\ell)} (1 - p_{\ell k}(s))^{\mathds{1}(s \in \ddot{z}_k)}\), $c_{\ell k} = \tilde{y}_k \bigr|_{\ddot{S}_\ell} \cup z_\ell \bigr|_{S_k \cap \ddot{S}_\ell}$ and $\ddot{c}_{\ell k} = \ddot{\tilde{y}}_k \bigr|_{S_\ell} \cup \ddot{z}_\ell \bigr|_{\ddot{S}_k \cap S_\ell}$. To compute the acceptance probability in Equation (\ref{eq__acceptance_probability__varphi}), it is necessary to retrospectively sample $\beta_\ell$ in a finite collection of locations, for $l = 1, \cdots, L$. The set of locations that require evaluation is typically small, as only the points near the boundaries of the regions tend to change label or region under the proposed partition.

A proof of the validity of this MH step, showing that it targets the correct full conditional distribution, is provided in the online Appendix A.4. Further details on the proposal mechanism are provided in the online Appendix B.1.

\subsubsection{Updating $\beta$}

The full conditional distribution of \(\beta\) is decomposed as:
\begin{equation}\label{eq__full_conditional__beta}
\pi(\beta \mid \cdot) \ \propto \
\pi(y, \tilde{y} \mid \lambda^*, \beta_{y^*}, u) \, 
\pi(\beta_{y^*} \mid \theta) \, 
\pi(\beta_{\mathcal{S} \setminus y^*} \mid \beta_{y^*}, \theta).
\end{equation}

The first term on the right-hand side of \eqref{eq__full_conditional__beta} corresponds to the product of the Poisson process likelihoods for the $Y_\ell$ and $\tilde{Y}_\ell$ processes. The second term is the Lebesgue density of $\beta_{y^*}$ under the GP prior, where $\beta_{y^*}$ denotes the combined vector of each $\beta_\ell$ at locations $y^*_\ell = (y_\ell,\tilde{y}_\ell)$. The third term represents the conditional GP prior for the infinite-dimensional remainder $\beta_{\mathcal{S} \setminus y^*}$, given $\beta_{y^*}$.

The component $\beta_{y^*}$ is sampled from a distribution whose density with respect to the Lebesgue measure is proportional to the product of the first two terms. When $F = \Phi$, this distribution belongs to the class of multivariate skew-normal distributions and can be sampled using the algorithm introduced by \cite{gonccalves2018exact, gonccalves2023corrigendum}. Alternative sampling strategies include data augmentation schemes such as those of \cite{albert1993bayesian} for the probit case ($F = \Phi$), and \cite{polson2013} for the logistic link function. Another option, which does not rely on data augmentation, is to update $\beta$ using a Metropolis-Hastings step, as described in the online Appendix B.2.

The infinite-dimensional remainder of $\beta$, denoted $\beta_{\mathcal{S} \setminus y^*}$, is sampled retrospectively from the corresponding conditional GP prior. This is done only at the finite set of locations required during the updates of the $(\tilde{Y}, Z)$ and $(U, Y, \tilde{Y}, Z)$ blocks. 

\subsubsection{Updating $\lambda^*$}

Conditionally independent Gamma full conditionals are obtained for \(\lambda_\ell^*\), given conjugate independent Gamma priors.

\subsubsection{Updating $(\theta,\beta_{\mathcal{S} \setminus y^*})$} \label{sec__updating_beta}

The full conditional distribution of $(\theta, \beta_{\mathcal{S} \setminus y^*})$ is decomposed as:
\begin{eqnarray}
\label{eq__full_conditional__theta_beta}
\pi(\theta, \beta_{\mathcal{S} \setminus y^*} \mid \cdot) &=& 
\pi(\theta \mid \cdot) \pi(\beta_{\mathcal{S} \setminus y^*} \mid \theta, \cdot) \propto 
\pi(\beta_{y^*} \mid \theta) \, 
\pi(\beta_{\mathcal{S} \setminus y^*} \mid \beta_{y^*}, \theta) \, \pi(\theta),
\end{eqnarray}
The marginal full conditional density of $\theta$ is proportional to $\pi(\beta_{y^*} \mid \theta) \, \pi(\theta)$ and typically requires a Metropolis–Hastings step, particularly for parameters that index the correlation function of the GP prior. A standard Gaussian random walk proposal generally performs well in this context. The proposal for $\beta_{\mathcal{S} \setminus y^*}$ is precisely $\pi(\beta_{\mathcal{S} \setminus y^*} \mid \beta_{y^*}, \theta)$, therefore leading to the acceptance probability
\begin{equation} \label{eq__acceptance_theta}
1 \ \wedge \ \frac{\pi(\beta_{y^*} \mid \ddot{\theta}) \, \pi(\ddot{\theta})}{\pi(\beta_{y^*} \mid \theta) \, \pi(\theta)}.
\end{equation}
The component $\beta_{\mathcal{S} \setminus y^*}$ is included in this block only to enable valid updates of the hyperparameter $\theta$. The proposed $\beta_{\mathcal{S} \setminus y^*}$ is only theoretically sampled (from the corresponding conditional GP prior) because the acceptance probability in \eqref{eq__acceptance_theta} does not depend on it.

\subsection{Further Computational Aspects}

The adopted MCMC framework facilitates the estimation of arbitrary functions of $\psi$ using posterior samples. For a function of interest $h(\psi)$, one simply evaluates $h(\psi)$ at each MCMC draw to obtain Monte Carlo estimates. Moreover, some intractable functions $h$ can be estimated without discretization error by employing unbiased estimators $\tilde{h}(\psi, u)$, where $u$ is an auxiliary variable.

For instance, to estimate $h(\psi) = \int_{A} \lambda(s;\psi)\,ds$, for some measurable subset $A \subset \mathcal{S}$, one can use the fact that
\[
\tilde{h}(\psi, u) \ = \ |A| \, \lambda(u;\psi), \quad u \sim U(A),
\]
is an unbiased estimator of $h(\psi)$. Variance reduction can be achieved by partitioning $A$ and sampling one uniform location per subregion. Posterior samples of $\tilde{h}(\psi,u)$ can then be used to approximate $h(\psi)$.

For spatial domains $\mathcal{S} \subset \mathbb{R}^d$ with $d \leq 2$, graphical representations of point estimates and pointwise credible intervals for the IF can be produced by evaluating the posterior of the IF on a fine mesh. At each MCMC iteration, this is done by drawing from the conditional distribution $\beta_{\mathcal{S} \setminus y^*} \mid \beta_{y^*}, \theta$.

A well-known computational bottleneck in Gaussian process models is the cubic cost associated with sampling from finite-dimensional multivariate Gaussian distributions. In the proposed model, this cost is substantially mitigated by the random partitioning approach, especially when compared to the special case with $L = 1$. 
The computational cost associated to the latent GP components $\{\beta_\ell\}$ scales as
$
\mathcal{O}\left(\sum_{\ell=1}^L (w_\ell n)^3\right),
$
where $n$ is the dataset size, $w_\ell > 0$ and $\sum_{\ell=1}^L w_\ell = 1$. This expression simplifies further to
\[
\mathcal{O}\left( \frac{n^3}{L^{2}} \right),
\]
when the weights $w_\ell \approx 1/L, \forall \ell$, therefore implying that, for a fixed $n$, the cost scales as $1/L^2$.

For larger datasets where the computational cost becomes prohibitive, the piecewise continuous GP (PCGP) of \cite{goncalves2025pcnngp} can be employed. This approach approximates the full GP through local conditional independence and spatial partitioning via a Vecchia-type construction. Crucially, it defines a valid GP prior measure and guarantees a measurable likelihood, yielding a well-defined posterior. The overall algorithm therefore remains exact in the sense that all sources of error are purely Monte Carlo. Alternatively, other Vecchia-type approximations such as the popular nearest-neighbor Gaussian process (NNGP) of \cite{datta2016hierarchical} can be used but are avoided here as they compromise exactness.

\section{Numerical Examples} \label{sec__applications}

The proposed methodology, hereafter referred to as the NSPP model (Nonstationary Point Process), is applied to both synthetic and real datasets to assess its performance. Results from two representative synthetic experiments are presented in Section \ref{sec__simulations}, with additional examples provided in the online Appendix C. Applications to two real datasets are discussed in Section \ref{sec__real_applications}. All analyses were implemented in the \texttt{R} programming language (version 4.2.2; \citealp{R2025}), with C++ integration via the \texttt{Rcpp} package (version 1.0.9; \citealp{rcpp2011}). All computations were performed on a Ubuntu 22.04.5 LTS machine with an 13th Gen Intel(R) Core(TM) i9-13900HX processor and 62 GB of RAM. Convergence is assessed empirically for all analyses and details are provided in the online Appendix E.

In all examples, the Gaussian process priors $\beta_\ell$ are specified with mean zero, variance $\sigma^2 = 4$, and the correlation function given in \eqref{eq__correlation_function}, with $\gamma = 1.9$ and $\phi$ chosen to reflect the spatial scale of the domain. Independent, weakly informative Gamma priors with shape and rate parameters equal to 0.001 are assigned to the parameters $\lambda_\ell^*$. For the repulsive prior on the partition locations $U$ in (\ref{eq__U_prior}), hyperparameters are set to $\eta = 1.5$ and $\nu = 4$. The proposal radius is tuned during an adaptive phase, in which it is adjusted every 1,000 iterations, to achieve appropriate acceptance rates in line with the optimal scaling results, with a larger radius set to twice the smaller radius and assigned weight 0.05. The number $b$ of neighbors updated at each step is set to approximately $\log(L)$. 

\subsection{Synthetic Data Examples} \label{sec__simulations}

All data examples are simulated from model (\ref{eq__model__1})--(\ref{eq__model__2}) over the domain $\mathcal{S} = [0, 10] \times [0, 10]$. A series of experiments was carried out, comprising scenarios with medium (up to around two thousand observations) and large datasets (where standard GP computations become prohibitive; \citealp{datta2016hierarchical}).

\subsubsection{Example 1: Discontinuous IF}

Example 1 considers the case $L=2$, with $\lambda^* = (5, 15)$ and true range parameters $\phi = (2.5, 0.5)$. The dataset contains 503 occurrences. The NSPP model is fit to this dataset using different values of $L$ to assess model robustness.

For comparison, the kernel intensity estimator of \cite{diggle1985kernel} and the stationary model of \cite{gonccalves2018exact, gonccalves2023corrigendum}, that corresponds to the NSPP with $L = 1$, are fitted. The kernel estimator uses the \texttt{R} function \texttt{bw.diggle} (from the package \texttt{spatstat.geom}) to select the bandwidth, and the function \texttt{as.ppp} (from the package \texttt{spatstat.explore}) to compute the estimator.
An empirical performance indicator can be provided by the mean square error $\frac{1}{M} \sum_{j=1}^M \left[ \lambda(s_j) - \hat{\lambda}(s_j) \right]^2$ of the IF over a mesh of $M$ locations in $\mathcal{S}$, where $\hat{\lambda}$ is an estimator of $\lambda$. In all synthetic examples, $M = 5{,}625$ and $\hat{\lambda}$ is the posterior mean of $\lambda$.

\begin{figure}[!ht]
    \centering
    \includegraphics[width=\linewidth]{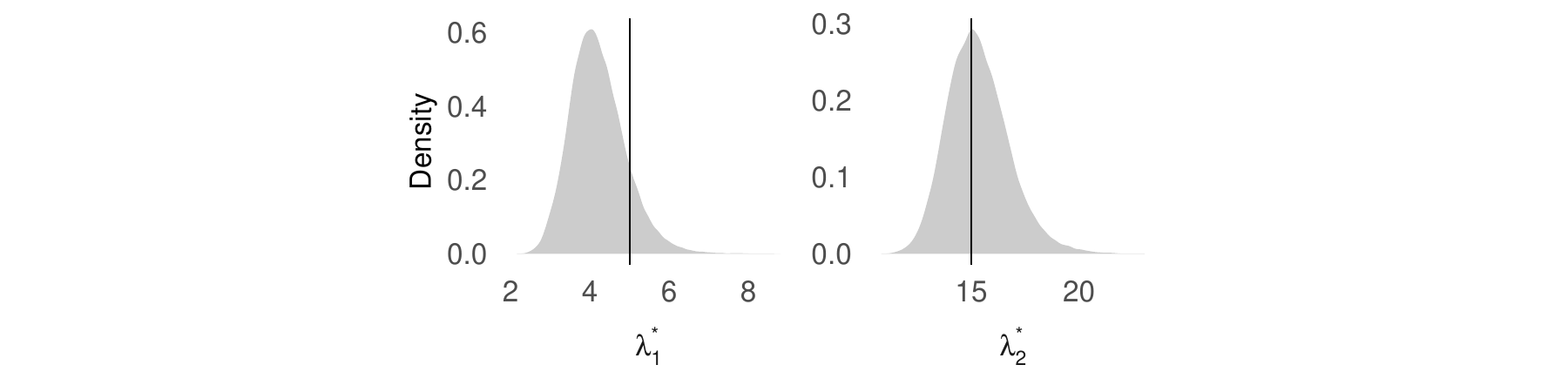}
    \caption{Marginal posterior densities of the $\lambda_{\ell}^*$ parameters in Example 1 for the model with $L = 2$. True value is indicated by the vertical line.}
    \label{fig__L2__lambdastar}
\end{figure}

\begin{figure}[!ht]
\centering
\includegraphics[width=\linewidth]{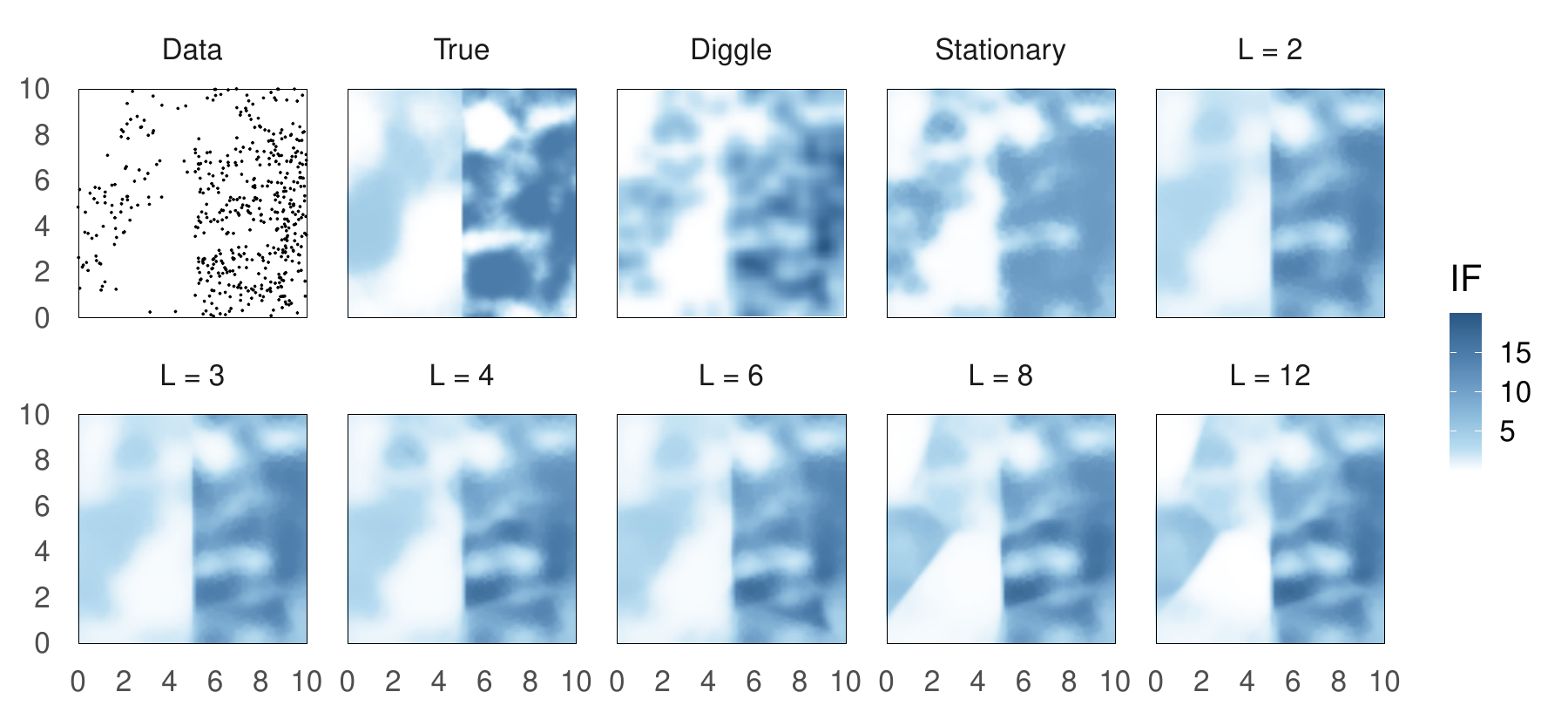}
\caption{Estimation of the IF in Example 1. The true IF has $L=2$ with a vertical boundary at $x=5$. The NSPP (for various $L$) is compared to the kernel estimator \citep{diggle1985kernel} and the stationary model ($L=1$).}
\label{fig__L2__if}
\end{figure}

Figure \ref{fig__L2__if} shows the posterior mean of the IF for different values of $L$. For all $L \geq 2$, the NSPP outperforms both the stationary and kernel-based models. The smallest value of the performance indicator is obtained with $L = 2$, with percentage growth of around 10\% for larger values of $L$. In contrast, stationary and kernel estimators yield larger increases of around 50\%. The stationary and kernel-based models notably fail to acknowledge the break in the IF, underestimate the intensity on the right half of $\mathcal{S}$ and overestimate it on the left. Figure \ref{fig__L2__lambdastar} displays the posterior marginals of the $\lambda_\ell^*$ parameters for $L = 2$.

\begin{figure}[!ht]
\centering
\includegraphics[width=\linewidth]{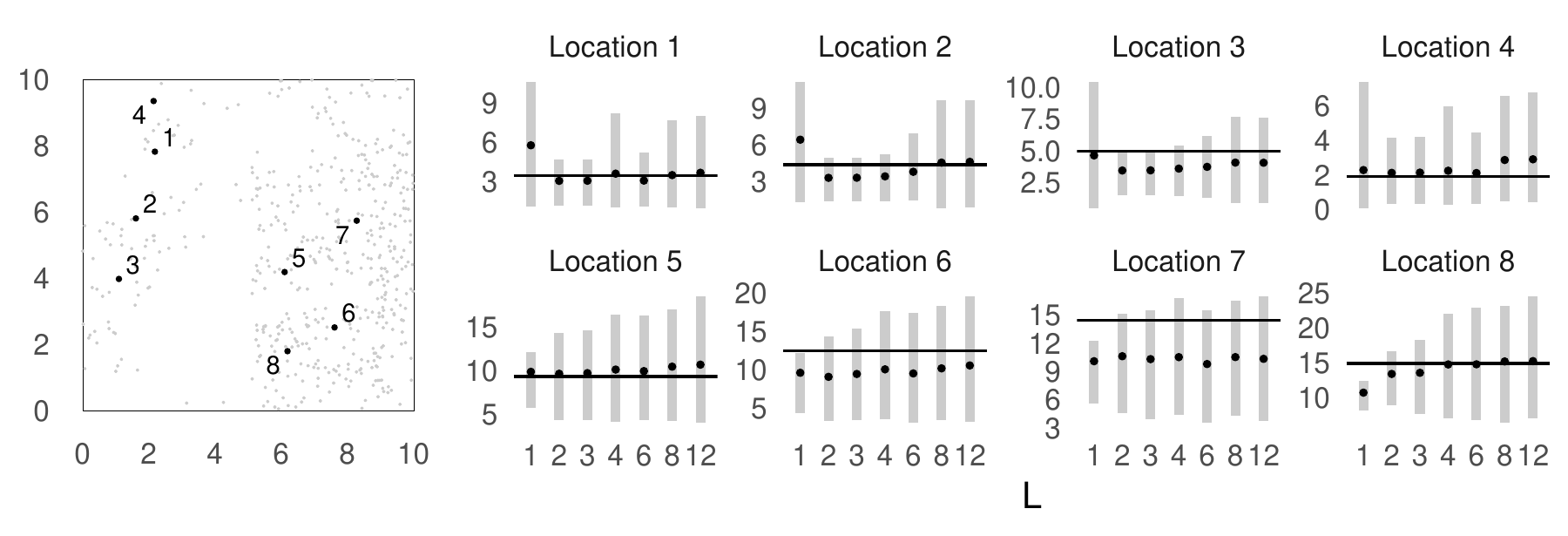}
\caption{Posterior estimation of the IF at selected locations in Example 1. Left: spatial layout. Right: posterior means (black dots), true values (black lines), and 95\% credibility intervals (gray bars) for different values of $L$.}
\label{fig__L2__add_loc}
\end{figure}

The posterior distribution of the IF was also examined at selected locations in Figure \ref{fig__L2__add_loc}. For all $L \geq 2$, the posterior concentration is close to the true intensity. In contrast, the $L = 1$ model yields a higher uncertainty on the left and a systematic underestimation on the right, except at location 5.

\begin{figure}[!ht]
\centering
\includegraphics[width=\linewidth]{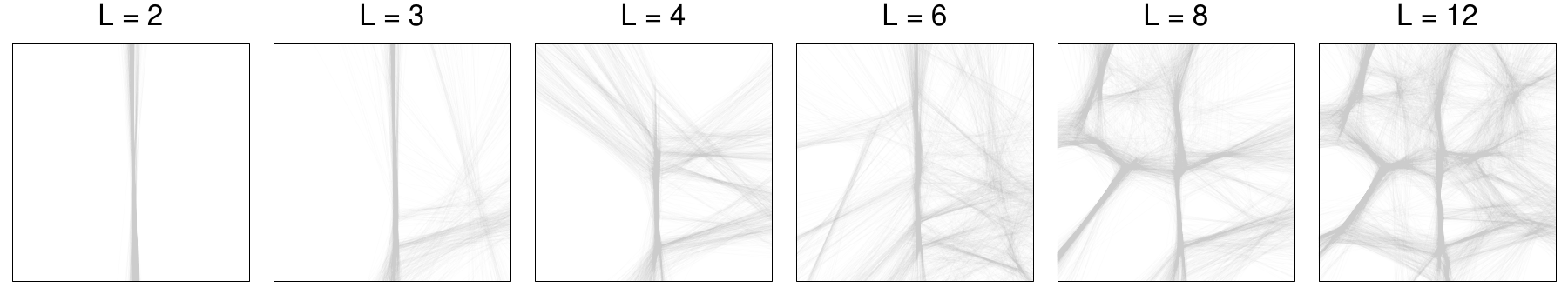}
\caption{Posterior samples of the partition for different values of $L$.}
\label{fig__L2__regions}
\end{figure}

Figure \ref{fig__L2__regions} shows 500 MCMC samples of the partition structure. As expected, only $L = 2$ matches the true configuration, but the discontinuity at $x=5$ is consistently detected at all values of $L$.

The posterior distribution of the partition also provides a diagnostic for structural breaks: boundaries supported by the data are estimated with high concentration, while computationally-driven, artificial boundaries exhibit high posterior uncertainty, arising when $L$ exceeds its true value. This feature allows for distinguishing discontinuities in the intensity surface from artifacts of model overspecification.

These results demonstrate the robustness of the NSPP model with respect to the choice of $L$. Accurate estimation of the IF is achieved even when $L$ is considerably larger than the true value, since the partition regions are large enough to provide information for useful inference. As expected, increasing $L$ leads to greater posterior uncertainty due to the conditional independence imposed across regions, as seen in the wider credible intervals in Figure \ref{fig__L2__add_loc}. In general, choosing moderately large values of $L$ introduces additional flexibility at a reasonable cost in uncertainty. Importantly, the primary goal of the NSPP is to provide flexible IF estimation rather than to recover the true partition structure. The observed robustness with respect to $L$ is especially appealing in real-data applications, where the true value of $L$ is unknown.

\subsubsection{Example 2: Hotspots}

\begin{figure}[!ht]
\centering
\includegraphics[width=\linewidth]{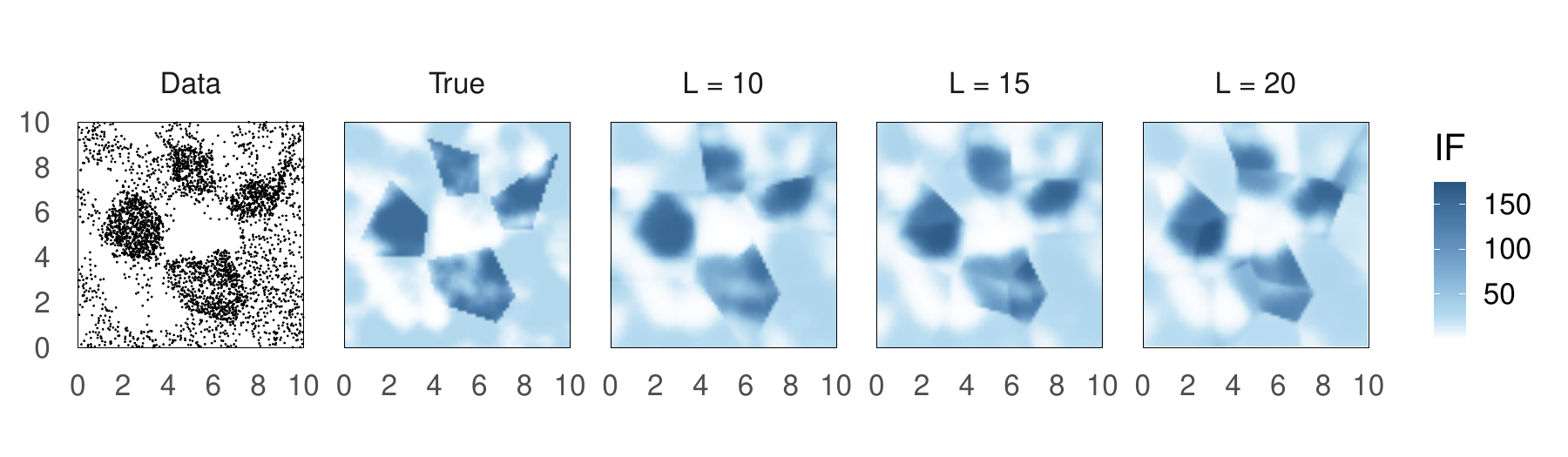}
\caption{Estimation of the IF in Example 2: data (dots), true and estimated IFs for different values of $L$. The true model has $L=14$ with localized hotspots.}
\label{fig__L14__if}
\end{figure}

A more challenging setting is considered in Example 2, where the true IF exhibits strong heterogeneity, with a few small regions of significantly elevated intensity, usually referred to as hotspots. Data are simulated with $L = 14$, with four regions having $\lambda_\ell^* = 150$ and the rest $\lambda_\ell^* = 30$, and $\phi=0.5$ for all regions. The dataset contains 3,742 occurrences. The NSPP is fitted using $L = 10$, $15$, and $20$ to assess sensitivity to the model specification. Figure \ref{fig__L14__if} shows the estimated IFs for each case. In all three scenarios, the IF is well recovered, indicating that the NSPP effectively adapts to localized spatial structure, even when the number of partitions is not correctly specified.

This example illustrates the ability of the NSPP to resolve localized high-intensity regions embedded within a low-intensity background. Notably, when $L$ is underspecified, the partition allocates regions preferentially around hotspots rather than distributing them uniformly, demonstrating the adaptation of the model to the most salient features of the spatial structure.

\subsubsection{Summary of Simulation Results}

The synthetic data examples demonstrate that the proposed NSPP methodology yields accurate and robust estimates of spatial IFs, even under model misspecification in the number of partitions $L$. In particular, the method performs well in both moderate and high-heterogeneity scenarios. Example 3 in online Appendix C shows that, even when the data are generated under $L=1$, the proposed model may outperform the stationary specification for larger values of $L$. This reinforces the flexibility of the nonstationary formulation and its ability to capture localized structure with greater accuracy. Additional experiments reported in the online Appendix C further support these findings.

\subsection{Real Data Analyses} \label{sec__real_applications}

\subsubsection{\textit{Beilschmiedia pendula} Data}

This dataset records the locations of 3,605 individuals of the species \emph{Beilschmiedia pendula} within a $1000\times500$ meter region on Barro Colorado Island, Panama \citep{hubbell1983diversity}. It has been previously analyzed by \cite{moller1998log} and \cite{hildeman2018level}, and is available through the \texttt{spatstat} package in \texttt{R} \citep{baddeley2015spatial}. For comparison, the log-Gaussian Cox process (LGCP; \citealp{moller2007modern}) and the nonstationary model (XGBoostPP; \citealp{Lu06082025}) are fitted using the \texttt{lgcp} \texttt{R} package \citep{lgcppackage} and the implementation at \url{github.com/Bobby-Lu/XGBoostPP}, respectively.

\begin{figure}[!ht]
\centering
\includegraphics[width=\linewidth]{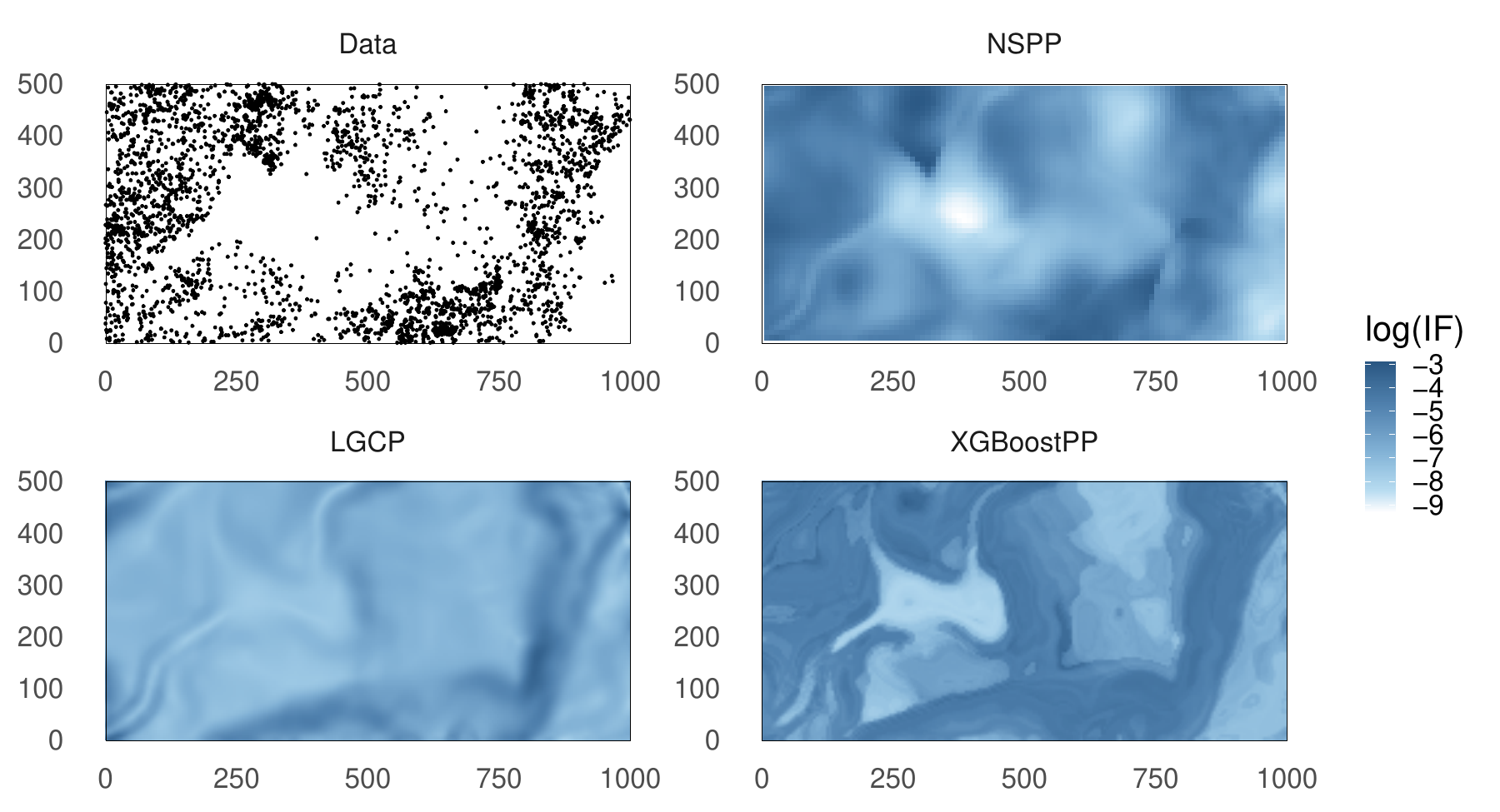}
\caption{Estimated log IF for the \textit{Beilschmiedia pendula} dataset. The posterior mean intensity under the NSPP with $L = 10$ is compared to the LGCP \citep{moller2007modern} and XGBoostPP \citep{Lu06082025}. All models include elevation and elevation gradient as covariates.}
\label{fig__bei__if}
\end{figure}

Covariate effects are incorporated into the NSPP model using globally shared coefficients across all partition regions. The model is fitted with $L = 10$, using elevation and the norm of the elevation gradient as covariates. The estimated IF, shown in Figure \ref{fig__bei__if}, captures both high- and low-density regions, consistent with the observed data pattern and indicating a good overall fit. Notably, the near-absence of trees in the central portion of the plot is reflected in a correspondingly low estimated intensity. The NSPP achieves this naturally through its partition structure, without requiring any prior specification of a low-intensity class, in contrast to previous analyses of the same dataset \citep{hildeman2018level}.

Figure \ref{fig__bei__if} also presents the estimated IFs from two competitive alternatives. The LGCP produces an overly smooth surface that fails to capture the pronounced high-intensity regions in the domain, underestimating the spatial contrasts present in the data. XGBoostPP assigns high intensity to the central portion of the domain, where the observed point density is low, producing a highly irregular surface with unreliable local estimates. The NSPP avoids both shortcomings: it accurately captures both intensity peaks and near-absence zones.

The posterior summaries of the regression coefficients are presented in Table \ref{tab__bei__coef} and suggest positive associations with both covariates. The effect of the elevation gradient norm is stronger than that of the elevation itself. This is consistent with the interpretation that this species favors sloped terrain, which could reflect preferences related to soil drainage or light exposure. The estimates are similar to those from \cite{moller2007modern}. Comparison of the two approaches must be exercised with care, due to important differences between them on inference procedure, GP specifications and link functions.

\begin{table}[!ht]
    \caption{Estimation of the regression coefficients for the \textit{Beilschmiedia pendula} dataset: posterior mean and 95\% credibility intervals (in brackets).}\label{tab__bei__coef}
    \centering
    \begin{tabular}{ccc}
    \hline
       Covariate  & \cite{moller2007modern} & NSPP\\ \hline
       Elevation in meters  &  0.06 [0.02, 0.10] & 0.08 [0.03, 0.13]\\
       Norm of elevation gradient  &  8.76 [6.03, 11.37]  & 8.06 [4.80,  13.70]   \\ \hline
    \end{tabular}
\end{table}

\begin{figure}[!ht]
\centering
\includegraphics[width=\linewidth]{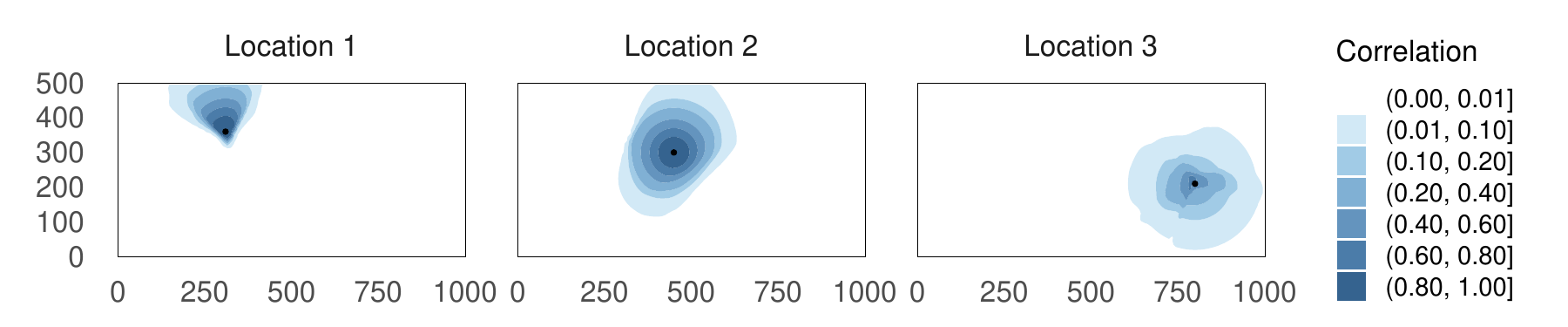}
\caption{Estimated correlation structure at three representative locations in the \textit{Beilschmiedia pendula} dataset.}
\label{fig__bei__corr}
\end{figure}

Figure \ref{fig__bei__corr} presents estimated spatial correlation maps at three representative locations. The results highlight the flexibility of the NSPP in capturing nonstationary spatial dependence. Location 2, situated well within a single partition region, displays smooth and approximately isotropic correlations. Locations 1 and 3, near partition boundaries, exhibit anisotropic patterns with weaker dependence in the direction crossing the boundary, consistent with a discontinuity in habitat conditions at these sites.

\subsubsection{Mato Grosso State Fires}

Mato Grosso is a Brazilian state in the Central-West region, covering approximately 903,207km$^2$ and home to around 3.6 million people. It borders the Amazon rainforest to the north and has experienced substantial agricultural expansion in recent decades. The dataset analyzed here consists of 8,177 fire occurrences detected during the 2023 dry season (May to August), obtained from the BDQueimadas database \citep{bdqueimadas2025} and the spatial boundaries provided by the \texttt{geobr} package \citep{geobr2024}.

The objective of this analysis is to characterize the spatial distribution of fire occurrences and identify potential high-risk regions, which are of direct relevance for fire prevention and land management. No covariates are included in this analysis.

\begin{figure}[!ht]
\centering
\includegraphics[width=\linewidth]{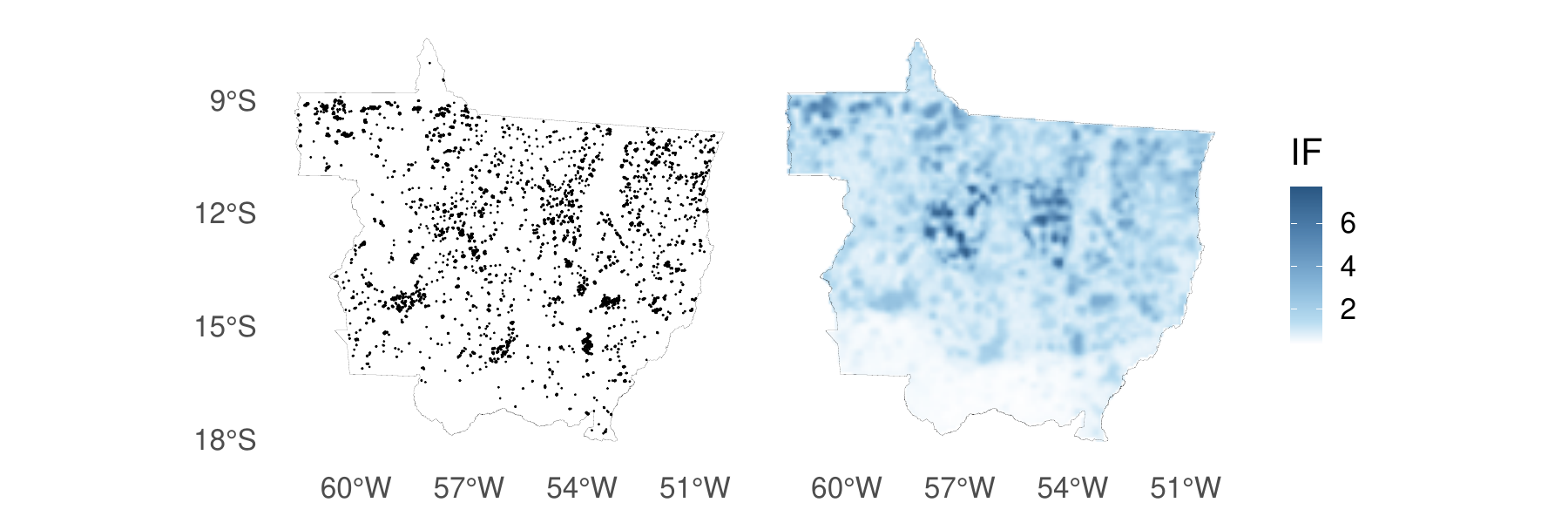}
\caption{Estimated IF for the Mato Grosso fires dataset: observed fire occurrences (dots) and posterior mean intensity under the NSPP model with $L = 20$.}
\label{fig__fires__if}
\end{figure}

The NSPP model is fitted with $L = 20$ partition regions. The estimated IF in Figure \ref{fig__fires__if} provides a smoothed representation of the underlying fire intensity and aligns well with the observed fire occurrences. Higher intensities are concentrated in the central-northern and northwestern parts of the state, while the southern region shows substantially lower activity.

\begin{figure}[!ht]
\centering
\includegraphics[width=\linewidth]{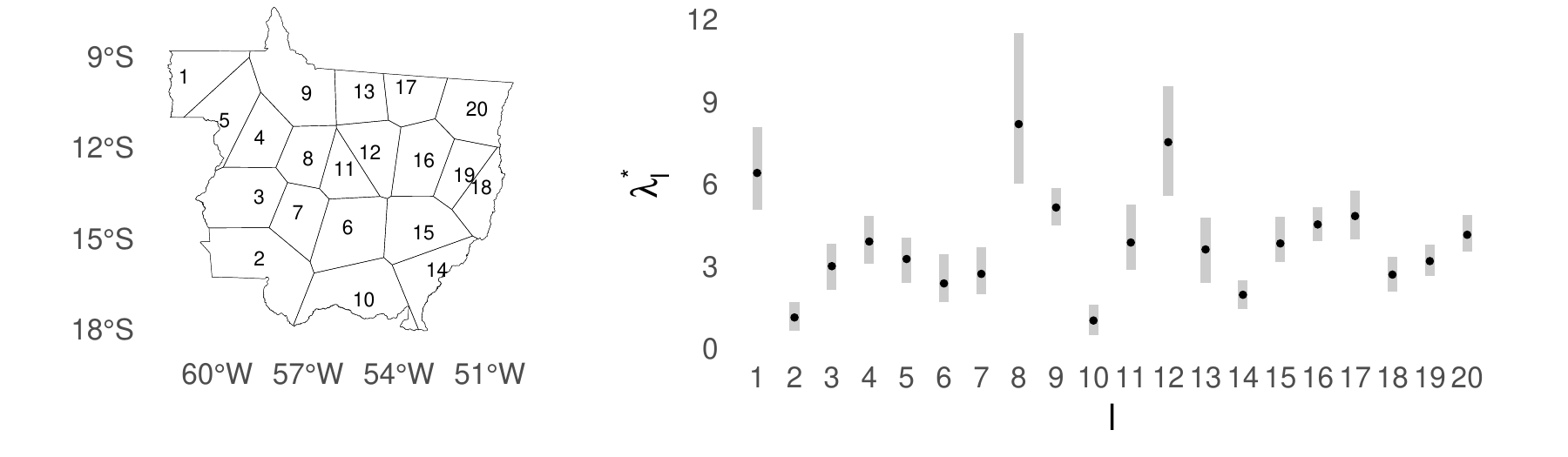}
\caption{Posterior summaries of the $\lambda_{\ell}^*$ parameters for the Mato Grosso fires dataset. Left: point estimate of the spatial partition based on the posterior mean of generating points. Right: posterior means (dots) and 95\% credibility intervals (gray bars) for each region.}
\label{fig__fires__lambdastar}
\end{figure}

Figure \ref{fig__fires__lambdastar} summarizes the posterior estimates of the parameters $\lambda_{\ell}^*$, which serve as upper bounds for the IF in each region. The highest intensity levels are found in regions 8 and 12 (both located in the north-central part of the state), followed by region 1 in the northwest. In contrast, southern regions -- such as 2, 10, and 14 -- exhibit significantly lower values, reflecting reduced fire incidence.

These findings are consistent with known land use dynamics in Mato Grosso: the northern half of the state overlaps with the arc of deforestation, where agricultural expansion is strongly associated with elevated fire incidence. The estimated partition identifies specific subregions of highest fire risk, providing spatially resolved information of direct relevance for fire prevention and environmental monitoring.

\section{Concluding Remarks}

This work introduced a nonparametric Bayesian model for nonstationary spatial point processes, along with a discretization-free MCMC algorithm for posterior inference. By using conditionally independent Gaussian processes over regions defined by a random partition, the proposed NSPP model captures complex and heterogeneous spatial structures, including sharp discontinuities. Its flexibility allows it to perform well even in stationary settings, outperforming stationary models in empirical comparisons.

Simulation studies demonstrated that the NSPP model accurately recovers intricate intensity surfaces and remains robust to overspecification of the number of partitions, consistently delivering competitive performance across a range of configurations. Compared to stationary, kernel-based, and state-of-the-art nonstationary approaches, the model achieved lower error rates and better accommodation of spatial discontinuities. Applications to real datasets further highlighted the model's interpretability and its ability to adapt to real-world spatial complexity in ecological and environmental contexts.

Future research directions include the incorporation of nonspatial covariates, which are particularly relevant in application areas such as economics and epidemiology. In these fields, spatial point process models often rely on individual-level covariates -- such as age, marital status, and education level -- that play a key role in explaining occurrences. Accommodating such covariates may require modifications to the current model structure. Improvements in computational efficiency are an important direction for future work.

Overall, the proposed methodology provides a flexible and principled framework for modern spatial point pattern analysis. The development of user-friendly software for practitioners is also underway and is expected to facilitate broader application of the NSPP model in both academic and applied settings.

\section{Acknowledgment}

The authors thank Vinicius Peripato for assistance with the Mato Grosso fires data. This research is part of the doctoral studies of the first author in the graduate program in Statistics at UFRJ, under the supervision of the second and third authors. The first author also thanks the program for institutional support and Murabei Data Science for computing support. The authors would like to thank the AE and the referees for their substantive comments that led to a much improved version of the manuscript. The research of Izabel Nolau is partially supported by CAPES. Flávio Gonçalves and Dani Gamerman are supported by FAPEMIG (APQ-01837-22) and CNPq (grants 308536/2023-1 and 302929/2022-8, respectively). Dani Gamerman is also supported by CNPq grant INCT 406913/2022-6.

\clearpage

\appendix

\begin{center}

{\LARGE \bf Supplemental Online Material}

\end{center}

\section{Proofs} \label{appendix__proofs}

\subsection{Augmented likelihood derivation}

The augmented likelihood function is given by the joint density of $(\mathcal{Y}, Y, \tilde{Y}, Z)$. Given the conditional independence of the processes and the Poisson likelihood formula, it follows that
\begin{eqnarray*}
\mathscrr{l}(\lambda^*, \beta, S, y, \tilde{y}, z \, ; \mathscrr{y}) 
&\propto& 
\pi(\mathscrr{y} \mid y) \,
\pi(y \mid \lambda^*, \beta, S) \,
\pi(\tilde{y} \mid \lambda^*, \beta, S) \,
\pi(z \mid \lambda^*, S),
\end{eqnarray*}
where each component is given by
\begin{eqnarray}
\label{eq__aug_model__data}
\pi(\mathscrr{y} \mid y) &=& \mathds{1}\left(\mathscrr{y} = \bigcup_{\ell=1}^L y_\ell\right), \\
\label{eq__aug_model__y}
\pi(y \mid \lambda^*, \beta, S) &=& \prod_{\ell=1}^L \exp\Big\{-\int_{S_\ell} \lambda_\ell^* F(\beta_\ell(s)) ds \Big\} \prod_{s \in y_\ell} \lambda_\ell^* F(\beta_\ell(s)), \\
\label{eq__aug_model__ytilde}
\pi(\tilde{y} \mid \lambda^*, \beta, S) &=& \prod_{\ell=1}^L \exp\Big\{-\int_{S_\ell} \lambda_\ell^* [1 - F(\beta_\ell(s))] ds \Big\} \prod_{s \in \tilde{y}_\ell} \lambda_\ell^* [1 - F(\beta_\ell(s))], \\
\label{eq__aug_model__z}
\pi(z \mid \lambda^*, S) &=& \prod_{\ell=1}^L \exp\Big\{- \lambda_\ell^* |\mathcal{S} \setminus S_\ell| \Big\} \prod_{s \in z_\ell} \lambda_\ell^*.
\end{eqnarray}
Note that $\pi(\mathscrr{y} \mid y)$ is the probability mass function of a degenerate distribution that verifies whether the union of the occurrences of processes $Y_1, \cdots, Y_\ell$ are equal to the observed data $\mathscrr{y}$. Standard calculations lead to the tractable form in (7).

\subsection{Proposition 1}

\begin{eqnarray*}
Cov(\beta(s), \beta(s'))
&=& E\left[Cov(\beta(s), \beta(s') \mid U)\right] + Cov\left(E[\beta(s) \mid U], E[\beta(s') \mid U]\right)\\
&=& E\left[\sum_{\ell = 1}^L\sigma^2_\ell\rho(h)\mathds{1}(s \in S_\ell, s' \in S_\ell)\right]  \ + \ \\
&& \sum_{\ell = 1}^L\sum_{m = 1}^L\mu_\ell(s)\mu_m(s') Cov\left(\mathds{1}(s \in S_\ell), \mathds{1}(s' \in S_m)\right)\\
&=& \sum_{\ell = 1}^L\sigma^2_\ell\rho(h)P(s \in S_\ell, s' \in S_\ell) \ + \ \\
&& \sum_{\ell = 1}^L\sum_{m = 1}^L\mu_\ell(s)\mu_m(s')\left[P(s \in S_\ell, s' \in S_m) - P(s \in S_\ell)P(s' \in S_m)\right].
\end{eqnarray*}

\subsection{Proposition 2}
\begin{eqnarray*}
&& Cov(\lambda(s), \lambda(s') \mid \lambda^*)\\
&=& Cov\left(
\sum_{\ell = 1}^L\lambda^*_\ell F(\beta_\ell(s))\mathds{1}(s \in S_\ell), 
\sum_{m = 1}^L\lambda^*_mF(\beta_m(s'))\mathds{1}(s' \in S_m)
 \ \middle| \ \lambda^*\right)\\
&=& E\left[Cov\left(\sum_{\ell = 1}^L\lambda^*_\ell F(\beta_\ell(s))\mathds{1}(s \in S_\ell), \sum_{m = 1}^L\lambda^*_mF(\beta_m(s'))\mathds{1}(s' \in S_m)  \ \middle| \ \lambda^*, U\right) \ \middle| \ \lambda^*\right] \ + \\
&& Cov\left(E\left[\sum_{\ell = 1}^L\lambda^*_\ell F(\beta_\ell(s))\mathds{1}(s \in S_\ell)  \ \middle| \ \lambda^*, U\right], E\left[\sum_{m = 1}^L\lambda^*_mF(\beta_m(s'))\mathds{1}(s' \in S_m)  \ \middle| \ \lambda^*, U\right] \ \middle| \ \lambda^*\right)\\
&=&E\left[\sum_{\ell = 1}^L\lambda^{*2}_\ell Cov\left(F(\beta_\ell(s)), F(\beta_\ell(s'))  \ \middle| \ \lambda^*, U\right)\mathds{1}(s \in S_\ell)\mathds{1}(s' \in S_\ell) \ \middle| \ \lambda^*\right] \ + \\
&& Cov\left(\sum_{\ell = 1}^L\lambda^*_\ell E\left[F(\beta_\ell(s))  \ \middle| \ \lambda^*, U\right]\mathds{1}(s \in S_\ell), \sum_{m = 1}^L\lambda^*_mE\left[F(\beta_m(s'))  \ \middle| \ \lambda^*, U\right]\mathds{1}(s' \in S_m) \ \middle| \ \lambda^*\right).
\end{eqnarray*}

\subsection{Validity of the MH step for $(U, Y, \tilde{Y}, Z)$}

First, note that the particular form of the label switching mechanism in the proposal distribution ensures that any move with positive probability has positive probability in the reverse direction, thereby not violating reversibility.

The proposal density factorizes as
\begin{eqnarray}\label{eq__proposal__varphi0}
q(\ddot{u}, \ddot{y}, \ddot{\tilde{y}}, \ddot{z} \mid u, y, \tilde{y}, z)
\ = \ q(\ddot{\tilde{y}}, \ddot{z} \mid \ddot{u}, u, y, \tilde{y}, z) \
q(\ddot{y}\mid \ddot{u}, u, y, \tilde{y}, z) \ 
q(\ddot{u} \mid u, y, \tilde{y}, z),
\end{eqnarray}
with
\begin{eqnarray*}
q(\ddot{\tilde{y}}, \ddot{z} \mid \ddot{u}, u, y, \tilde{y}, z)
&=& q(\ddot{\tilde{y}}, \ddot{z} \mid \ddot{u}, \tilde{y}, z) \\
&=& \prod_{\ell = 1}^L q\left(\ddot{\tilde{y}}_\ell \, \Bigr| \, \ddot{u}, \, \tilde{y}_\ell \bigr|_{\ddot{S}_\ell} \right)
q\left(\ddot{z}_\ell \, \Bigr| \, \ddot{u}, \, z_\ell \bigr|_{\mathcal{S} \setminus \ddot{S}_\ell} \right)
q\left(\ddot{\tilde{y}}_\ell, \ddot{z}_\ell \, \Bigr| \, \ddot{u}, \, \tilde{y}_\ell \bigr|_{\mathcal{S} \setminus \ddot{S}_\ell}, \, z_\ell \bigr|_{\ddot{S}_\ell} \right),\\
q(\ddot{y}\mid \ddot{u}, u, y, \tilde{y}, z)
&=& q(\ddot{y}\mid \ddot{u}, y) \\
&=& \prod_{\ell = 1}^L \mathds{1}\left(\ddot{y}_\ell = \left[\bigcup_{\ell = 1}^L y_\ell\right] \bigr|_{\ddot{{S}}_\ell}\right),\\
q(\ddot{u} \mid u, y, \tilde{y}, z)
&=& q(\ddot{u} \mid u) \\
&=& 
\frac{1}
{L}\prod_{l \in \mathcal{N}_{b}(\mathscrr{l}^*)} \left[p \frac{||\ddot{u}_\ell - u_\ell||}{2 \pi r^2} + (1 - p)\frac{||\ddot{u}_\ell - u_\ell||}{2 \pi (mr)^2}\right],
\end{eqnarray*}
where $r$ is the smaller radius and $mr$ is the larger one, $q\left(\ddot{\tilde{y}}_\ell \, \Bigr| \, \ddot{u}, \, \tilde{y}_\ell \bigr|_{\ddot{S}_\ell} \right) = \mathds{1}\left(\ddot{\tilde{y}}_\ell \supseteq \tilde{y}_\ell \bigr|_{\ddot{S}_\ell} \right)
$,
$q\left(\ddot{z}_\ell \, \Bigr| \, \ddot{u}, \, z_\ell \bigr|_{\mathcal{S} \setminus \ddot{S}_\ell} \right) = \mathds{1}\left(\ddot{z}_\ell \supseteq z_\ell \bigr|_{\mathcal{S} \setminus \ddot{S}_\ell} \right)
$, and
\noindent $q\left(\ddot{\tilde{y}}_\ell, \ddot{z}_\ell \, \Bigr| \, \ddot{u}, \, \tilde{y}_\ell \bigr|_{\mathcal{S} \setminus \ddot{S}_\ell}, \, z_\ell \bigr|_{\ddot{S}_\ell} \right) = \prod_{\substack{ k = 1 \\ k \neq l}}^L q_{\ell k}(s)$, with $q_{\ell k}(s)$ being the probability mass function of a Bernoulli distribution with probability given in (16).

To obtain the expression of the acceptance probability of a move $u, y, \tilde{y}, z\rightarrow\ddot{u}, \ddot{y}, \ddot{\tilde{y}}, \ddot{z}$, define $\mathcal{X} = \bigcup_{\ell = 1}^L(Y_\ell \cup \tilde{Y}_\ell \cup Z_\ell)$. Standard Poisson process results imply that $\mathcal{X} \sim PP_\mathcal{S} (\lambda^*_+)$ where $\lambda^*_+ = \sum_{\ell = 1}^L \lambda^*_\ell$ and, for a fixed value $\mathscrr{x}$ of $\mathcal{X}$, the probability mass function $\pi(y, \tilde{y}, z \mid \mathscrr{x}, \lambda^*, \beta, u)$ is given by
\begin{equation}\label{eq__fc_uyyz}
\prod_{\ell = 1}^L \left[ \prod_{s \in y_\ell}\frac{\lambda^*_\ell F(\beta_\ell(s))}{\lambda^*_+}\prod_{s \in \tilde{y}_\ell}\frac{\lambda^*_\ell[1 - F(\beta_\ell(s))]}{\lambda^*_+} \prod_{\substack{ k = 1 \\ k \neq l}}^L \prod_{s \in z_{k}} \frac{\lambda^*_{k}}{\lambda^*_+}\mathds{1}(s \in S_{k})\right].
\end{equation}

Because $\mathcal{X}$ is fixed in this step, the dominating measure of the full conditional density and of the proposal distribution, at the current and proposal values, are the same, namely $\delta^N \otimes \mathbb{L}^{dL}$, where $\delta$ is the counting measure on $\mathds{N}$, $\mathbb{L}^{d}$ is the $d$-dimensional Lebesgue measure and $N = | \mathcal{X} |$. Therefore \citep[see][]{tierney1998note}, the acceptance probability is given by
\begin{equation}\label{eq__ap__uyyz}
1 \ \wedge \ \frac{\pi(\ddot{u}, \ddot{y}, \ddot{\tilde{y}}, \ddot{z} \mid \cdot \, ) \, q(u, y, \tilde{y}, z \mid \ddot{u}, \ddot{y}, \ddot{\tilde{y}}, \ddot{z})}{\pi(u, y, \tilde{y}, z \mid \cdot \, ) \, q(\ddot{u}, \ddot{y}, \ddot{\tilde{y}}, \ddot{z} \mid u, y, \tilde{y}, z)}.
\end{equation}

Finally, substituting (9), \eqref{eq__proposal__varphi0} and \eqref{eq__fc_uyyz} into \eqref{eq__ap__uyyz} and making suitable simplifications leads to the expression in (17).

\section{Details on the MCMC sampling}
\label{MCMCdetails}

\subsection{Updating $(U, Y, \tilde{Y}, Z)$} \label{appendix__sampling_varphi}

For the purpose of presentation clarity, assume, without loss of generality, that $\mathcal{S} \subset \mathbb{R}^2$. Representing $u_\ell$ by its Cartesian coordinates $(x_\ell, y_\ell)$, a proposed value $\ddot{u}_\ell = (\ddot{x}_\ell, \ddot{y}_\ell)$ is generated according to $\ddot{x}_\ell = x_\ell + \tau \cos \vartheta$ and $\ddot{y}_\ell = y_\ell + \tau \sin \vartheta$, where $\vartheta \sim U(0, 2\pi)$ and $\tau \sim r\sqrt{U(0, 1)}$, with $r$ being the radius of the circle. The induced proposal density of $U_\ell$ conditional on $r$ is
\[
q(\ddot{u}_\ell \mid u_\ell, r) 
\ = \ q(\tau(\ddot{x}_\ell, \ddot{y}_\ell) \mid x, y) \, q(\vartheta(\ddot{x}_\ell, \ddot{y}_\ell) \mid x, y) \, \left|\frac{\partial(\tau, \vartheta)}{\partial(\ddot{x}_\ell, \ddot{y}_\ell)}\right| \
= \ \frac{||\ddot{u}_\ell - u_\ell||}{2\pi r^2}.
\]

The allocation structure of $(\tilde{y}, z)$ into $(\ddot{\tilde{y}}, \ddot{z})$ is presented in Figure \ref{fig__alocation}.

\tikzstyle{bag_s} = [draw, text centered]
\tikzstyle{bag_c} = [circle, draw, text width = 0.5cm, text centered]
\begin{figure}[ht!]
    \scriptsize
	\centering
	\begin{tikzpicture}[sloped]
		\matrix[row sep=0.25cm, column sep = 0.25cm] (M) {%
			  \node[bag_s] (Y1) {$\tilde{y}_{1} \bigr|_{\ddot{S}_\ell}$};
			& \node[] (d) {$\cdots$};
			& \node[bag_s] (Ylm1) {$\tilde{y}_{l - 1} \bigr|_{\ddot{S}_\ell}$};
			&
			& \node[bag_s] (Yl) {$\tilde{y}_\ell \bigr|_{\ddot{S}_\ell}$};
			&
			& \node[bag_s] (Ylp1) {$\tilde{y}_{l + 1} \bigr|_{\ddot{S}_\ell}$};
			& \node[] (d) {$\cdots$};
			& \node[bag_s] (YL) {$\tilde{y}_{L} \bigr|_{\ddot{S}_\ell}$};
			\\ \ \\ \ \\
			  \node[bag_c] (RZ1) {$\ddot{z}_{1}$};
			& \node[] (d) {$\cdots$};
			& \node[bag_c] (RZlm1) {$\ddot{z}_{l - 1}$};
			&
			& \node[bag_c] (RYl) {$\ddot{\tilde{y}}_\ell$};
			&
			& \node[bag_c] (RZlp1) {$\ddot{z}_{l + 1}$};
			& \node[] (d) {$\cdots$};
			& \node[bag_c] (RZL) {$\ddot{z}_{L}$};
			&
			& \node[bag_c] (RZl) {$\ddot{z}_\ell$};
			\\ \ \\ \ \\
			  \node[bag_s] (ZS1Sl) {$z_\ell \bigr|_{S_1 \cap \ddot{S}_\ell}$};
			& \node[] (d) {$\cdots$};
			& \node[bag_s] (ZSlm1Sl) {$z_\ell \bigr|_{S_{l - 1} \cap \ddot{S}_\ell}$};
			& 
			&
			&
			& \node[bag_s] (ZSlp1Sl) {$z_\ell \bigr|_{S_{l + 1} \cap \ddot{S}_\ell}$};
			& \node[] (d) {$\cdots$};
			& \node[bag_s] (ZSLSl) {$z_\ell \bigr|_{S_\ell \cap \ddot{S}_\ell}$};
			&
			& \node[bag_s] (ZSml) {$z_\ell \bigr|_{\mathcal{S} \setminus \ddot{S}_\ell}$};\\
				};
				 \draw[->] (Y1) -- (RYl);
				 \draw[->] (Y1) -- (RZ1);
				 \draw[->] (Ylm1) -- (RYl);
				 \draw[->] (Ylm1) -- (RZlm1);	
				 \draw[->] (Yl) -- (RYl);
				 \draw[->] (Ylp1) -- (RYl);
				 \draw[->] (Ylp1) -- (RZlp1);
				 \draw[->] (YL) -- (RYl);
				 \draw[->] (YL) -- (RZL);
				 \draw[->] (Y1) -- (RYl);
\draw[->] (ZS1Sl) -- (RZ1);
\draw[->] (ZS1Sl) -- (RYl);
\draw[->] (ZSlm1Sl) -- (RZlm1);	
\draw[->] (ZSlm1Sl) -- (RYl);
\draw[->] (ZSlp1Sl) -- (RYl);
\draw[->] (ZSlp1Sl) -- (RZlp1);
\draw[->] (ZSLSl) -- (RYl);
\draw[->] (ZSLSl) -- (RZL);	
\draw[->] (ZSml) -- (RZl);
			\end{tikzpicture} \vspace{-0.2cm}
			\caption{Allocation possibilities for a fixed $l$.}\label{fig__alocation}
		\end{figure}

To compute the acceptance probability in equation (17), it is necessary to retrospectively sample $\beta$ at a finite collection of locations, specifically, at $y_\ell\bigr|_{\mathcal{S} \setminus \ddot{S}_\ell}$, $\tilde{y}_\ell\bigr|_{\mathcal{S} \setminus \ddot{S}l}$, and $z_\ell\bigr|_{\ddot{S}_\ell}$, for $l = 1, \ldots, L$. The collection is typically small, since only the points near the boundaries of $S$ tend to change region under the proposed partition.

\subsection{Updating $\beta$ via Metropolis-Hastings}
\label{appendix__updating_beta}

A noncentered random walk proposal is adopted for $\beta_\ell$, for $l = 1, \cdots, L$. Specifically, a move $\beta_\ell \to \ddot{\beta}_\ell$ is proposed from
\begin{eqnarray}
    \ddot{\beta}_\ell(s) &=& \sqrt{1 - \varsigma^2}\beta_\ell(s) + \varsigma \varepsilon_\ell(s), \quad \text{for } s \in \mathcal{S}\\
   \nonumber \varepsilon_\ell &\sim& \mbox{GP}(0, \Sigma_\ell),
\end{eqnarray}
where $\Sigma_\ell$ is the prior covariance function of $\beta_\ell$. This is called the preconditioned Crank Nicolson (pCN) proposal, introduced in \cite{cotter2013mcmc}, and is used for updating latent GP components in \cite{gonccalves2022beyond} and \cite{gonccalves2023exactlevelset}. A key property of
the pCN proposal is that it is reversible w.r.t. the prior GP measure and, therefore, the resulting acceptance probability is given by the likelihood ratio. Specifically, the acceptance probability of a move $\beta_\ell \to \ddot{\beta}_\ell$ is given by
\begin{eqnarray}
1 \wedge
\frac{\displaystyle \prod_{s \in y_\ell}F(\ddot{\beta}_\ell(s))\prod_{s \in \tilde{y}_\ell}[1 -F(\ddot{\beta}_\ell(s))]}
{\displaystyle \prod_{s \in y_\ell}F(\beta_\ell(s))\prod_{s \in \tilde{y}_\ell}[1 -F(\beta_\ell(s))]}.
\end{eqnarray}

\section{Other application results} \label{appendix__other_results}

This section presents the analysis of three additional simulated datasets. Example 3 considers data generated from a stationary model ($L = 1$), while Example 4 involves a more complex nonstationary configuration with $L = 10$, where the values of $\lambda_\ell^*$ are set to 24 in the bottom-left region, 16 in four other high-intensity regions, and 4 elsewhere. Example 5 considers a large-scale dataset, designed to assess the practical applicability of the methodology at scales substantially larger than those in the main examples.
 
\begin{figure}[!ht]
\centering
\includegraphics[width=\linewidth]{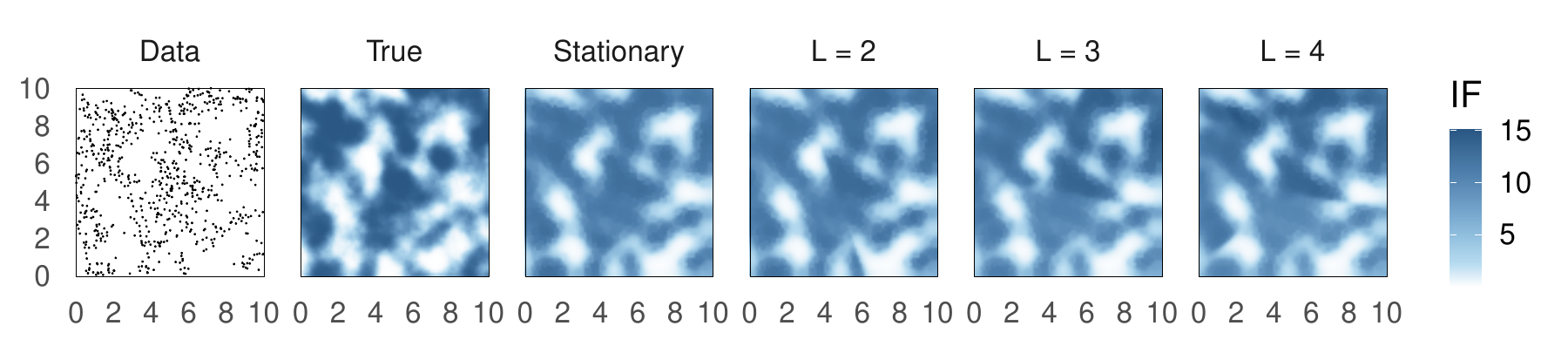}
\caption{Estimated IFs for Example 3: data (dots), true intensity, and posterior means under different values of $L$.}
\label{fig__L1__if}
\end{figure}

For Example 3, the NSPP model is fitted using four different values of $L$, including the true value $L = 1$. The dataset contains 755 occurrences. Figure \ref{fig__L1__if} shows the estimated IFs for each case. The results indicate that the true IF is accurately recovered even when $L$ is overspecified. The smallest value of the performance indicator is obtained with $L = 3$, with percentage growth of around 3\% for other values of $L$.

\begin{figure}[!ht]
\centering
\includegraphics[width=\linewidth]{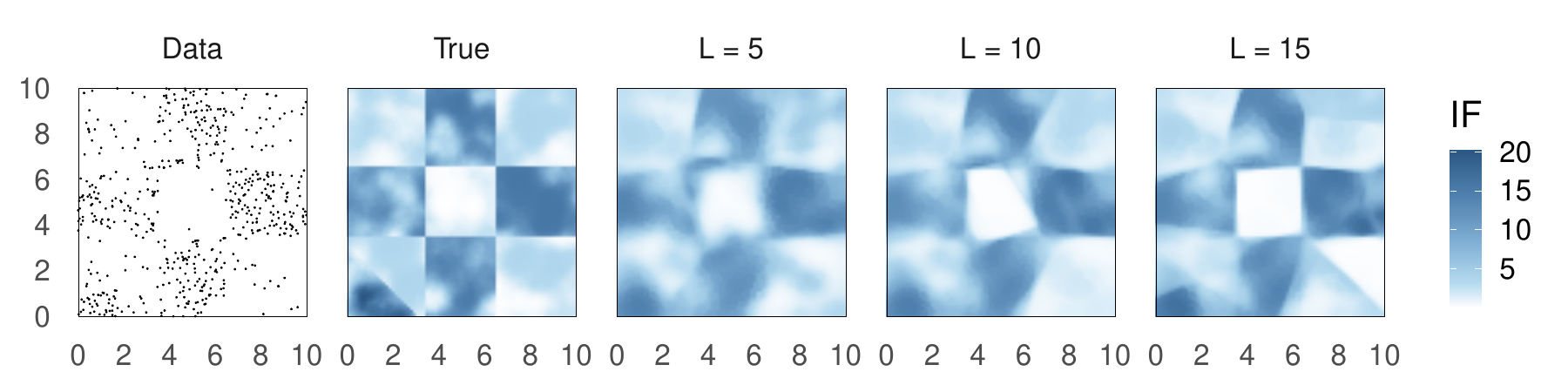}
\caption{Estimated IFs for Example 4: data (dots), true intensity, and posterior means under different values of $L$.}
\label{fig__L10__if}
\end{figure}

In Example 4, the dataset contains 562 occurrences. The NSPP model is fitted using three values of $L$, including the true value $L=10$. The corresponding estimated IFs are shown in Figure \ref{fig__L10__if}. The results demonstrate that the model continues to perform well in recovering spatial structure across multiple levels of heterogeneity.

\begin{figure}[!ht]
\centering
\includegraphics[width=\linewidth]{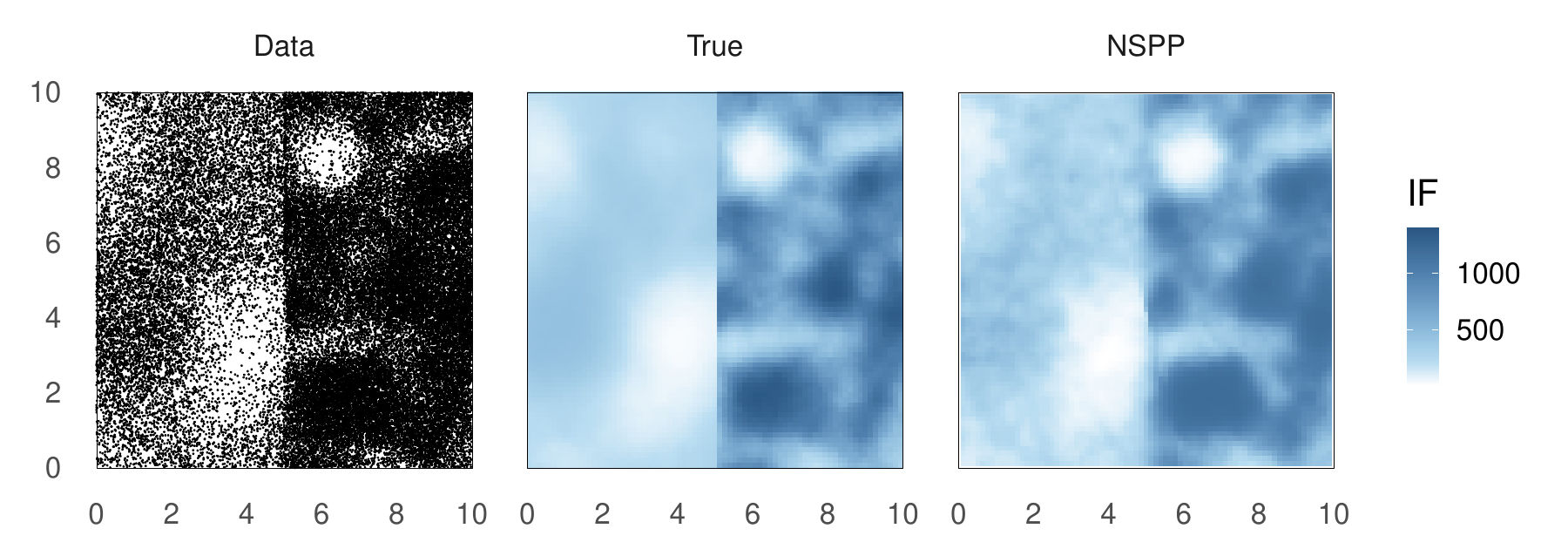}
\caption{Estimation of the IF in Example 5: observed data (dots), true intensity, and posterior mean under the NSPP model with $L = 2$.}
\label{fig__L2__large__if}
\end{figure}

In Example 5, the NSPP model with $L = 2$ is fitted to a dataset of $50{,}110$ occurrences simulated from the same configuration as Example 1, with $\lambda^* = (500, 1500)$. The estimated IF, shown in Figure \ref{fig__L2__large__if}, accurately recovers the true intensity surface and correctly identifies the discontinuity at $x = 5$, consistent with the results of Example 1.
 
To assess computational scalability, the runtime of the algorithm is evaluated across different sample sizes and numbers of partition regions. Two nested subdomains of the Example 5 dataset are considered: a medium-scale dataset comprising the region $[2.11, 7.89]^2$, which covers approximately one third of the total area and contains 15,177 occurrences, and a small-scale dataset comprising the region $[3.33, 6.67]^2$, which covers approximately one ninth of the total area and contains 5,073 occurrences. The full dataset is referred to as the large-scale dataset. The NSPP model is fitted to the small-scale dataset under $L = 2$, $5$, and $10$, and to the medium- and large-scale datasets under $L = 2$.

For $L = 2$, generating 10,000 iterations requires approximately 47.5 minutes for the small-scale dataset, corresponding to runtime increases of approximately 150\% and 780\% for the medium- and large-scale datasets, respectively, relative to the small-scale dataset. For fixed $n$ (small-scale dataset), increasing $L$ from 2 to 5 and 10 increases runtime by factors of approximately $1.9$ and $2.1$, respectively. These results indicate that runtime scales moderately with both $n$ and $L$, with the effect of $n$ being more pronounced.

\section{Guidances}

\subsection{Choice of hyperparameters $\eta$ and $\nu$}

Let $r(d) = 1 - \exp\{-\eta d^\nu\}$ denote the repulsive factor in Equation (9). This term acts as a soft regularization constraint: it penalizes partitions where centroids are excessively close ($r(d) \to 0$), while allowing data-supported configurations ($r(d) \to 1$) to remain unaffected.

To ensure the regularization is effective across different spatial domains, we recommend a two-step calibration process based on the domain's maximum extent, $D$ (e.g., the bounding box diagonal):

\begin{enumerate}
    \item \textbf{Shape parameter ($\nu$):} This controls the sharpness of the boundary between ``forbidden'' and ``allowed'' distances. 
    \begin{itemize}
        \item Lower values ($\nu \in [1, 2]$) create a gradual penalty that may unnecessarily affect centroids at moderate distances.
        \item Higher values ($\nu \in [3, 4, 5]$) produce a sharper ``step-like'' behavior, concentrating the repulsion only on very small separations. We recommend $\nu \approx 4$ as a robust default to enforce minimal separation without imposing hard constraints.
    \end{itemize}
    
    \item \textbf{Scale parameter ($\eta$):} This determines the active range of the repulsion. Since distances are relative, $\eta$ should be calibrated such that the repulsive effect is active only for a small fraction of $D$. A practical heuristic is to visualize $r(d)$ and select $\eta$ so that $r(d)$ approaches 1 as soon as $d$ exceeds a small tolerance threshold (e.g., $1\%$ to $2\%$ of $D$).
\end{enumerate}

\begin{figure}[!ht]
\centering
\includegraphics[width=\linewidth]{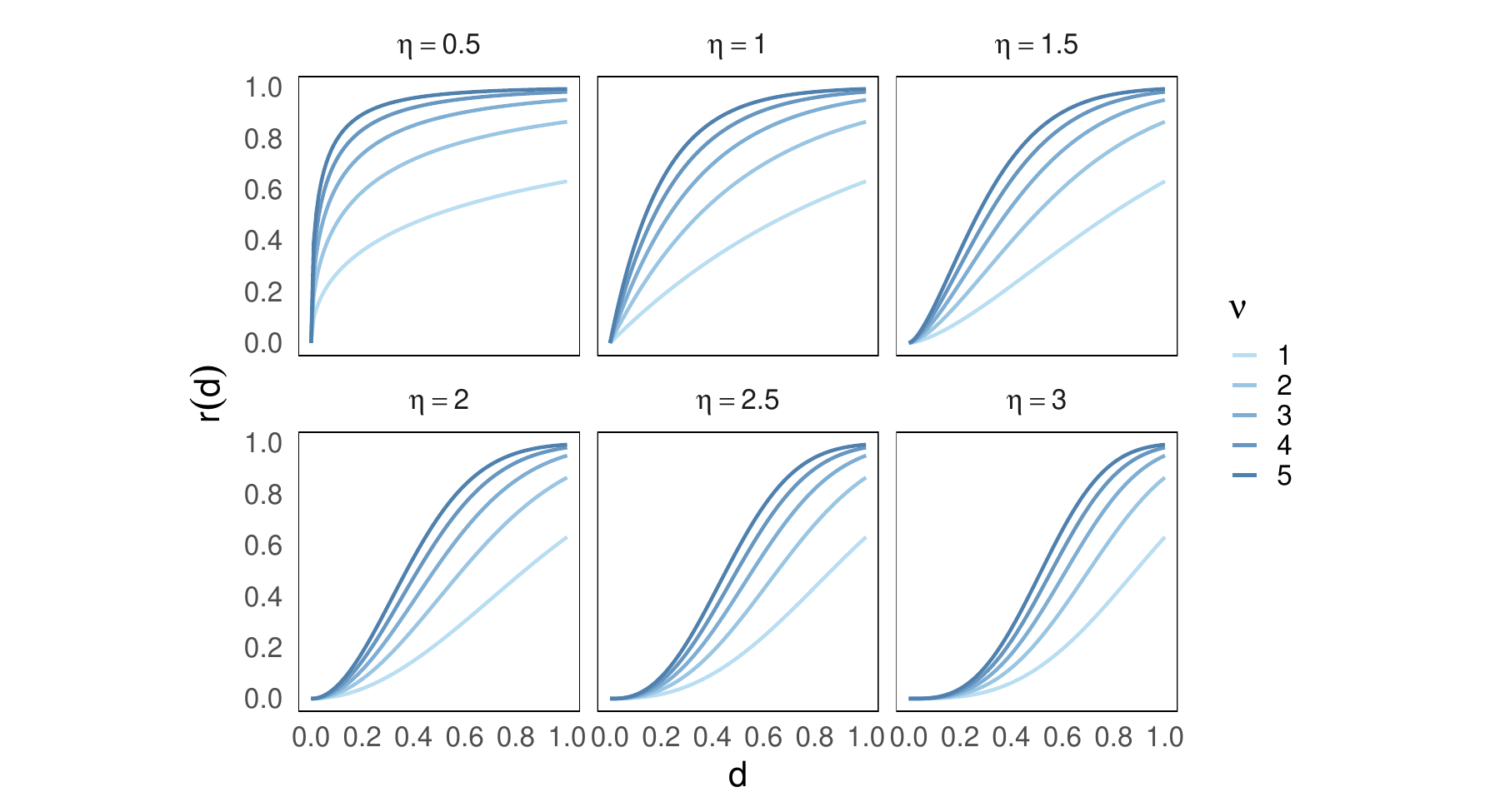}
\caption{Penalizing factor $r(d) = 1 - \exp \{-\eta \, d^\nu\}$ for different values of $\eta$ and $\nu$.}
\label{fig__repulsive_term}
\end{figure}

Figure \ref{fig__repulsive_term} serves as a calibration template. Although generated for a domain with $D \approx 14.14$, the curves are universally applicable if the horizontal axis is interpreted as relative distance. For the numerical examples in this paper ($\mathcal{S} = [0,10]^2$, $D \approx 14.14$), the pair $\eta = 1.5$ and $\nu = 4$ was selected. This configuration ensures that centroid separations below $\approx 0.5$ units are heavily penalized. For a different spatial domain, the same curves apply after rescaling distances by the corresponding maximum distance $D$. 

\subsection{Choice of the number of regions $L$}

The number of regions $L$ is treated as a user-specified quantity. In practice, the choice of $L$ should balance flexibility and estimability. Increasing $L$ allows for stronger spatial heterogeneity, at the cost of reduced information per region and potentially unstable estimation. Let $n$ denote the total number of observed events. On average, each region contains approximately $n/L$ occurrences. Stable estimation of the Gaussian processes requires this quantity to be sufficiently large, typically on the order of a few hundred observations \citep[see][]{webster1992sample}. For example, when $n=10{,}000$, values of $L$ up to approximately 20 lead to reliable estimation, while larger values may result in poorly identified regional effects.

\section{Assessment of MCMC convergence}\label{appendix__convergence}

Convergence of the MCMC algorithm was assessed for all analyses presented in the paper. For brevity, only a representative subset is displayed below. All remaining analyses indicated satisfactory convergence.

Consider the following function defined on a regular grid $\mathscrr{g} \subset \mathcal{S}$:
\begin{eqnarray}\label{eq__discretized_likelihood}
h(\lambda^*, \beta, S, y) &=& \prod_{\ell = 1}^L \left[\exp\left\{- \sum_{s \in \mathscrr{g}|_{S_\ell}} \lambda^*_\ell F(\beta_\ell(s))\right\} 
\prod_{s \in y_\ell} \lambda^*_\ell F(\beta_\ell(s))\right].
\end{eqnarray}

The expression in \eqref{eq__discretized_likelihood} corresponds to a discretized version of the likelihood in (3). The thinner the grid, the closer \eqref{eq__discretized_likelihood} becomes to the likelihood in (3). In this appendix, the term pseudo log-likelihood refers to the function defined in \eqref{eq__discretized_likelihood}, evaluated on samples recorded every 10 iterations.

\begin{figure}[ht!]
    \centering
    \includegraphics[width = \textwidth]{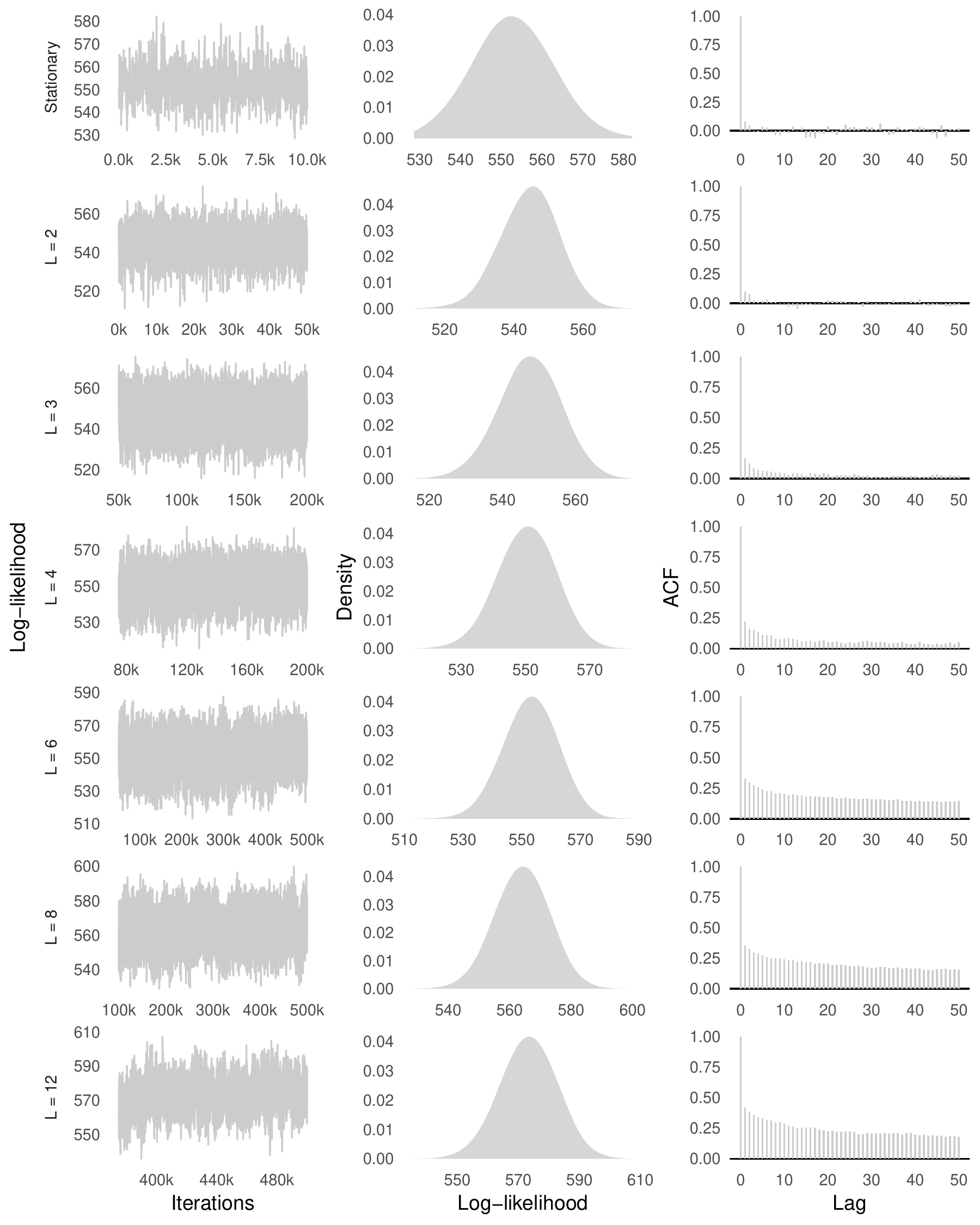}
    \caption{Trace plot, density estimates and autocorrelation function of the pseudo log-likelihood samples obtained for Example 1.}
    \label{fig__L2__convergence}
\end{figure}

Figure \ref{fig__L2__convergence} displays convergence diagnostics for Example 1. The ESS per 10,000 samples decreases with $L$: 8,458 (Stationary), 7,012 ($L=2$), 2,647 ($L=3$), 1,283 ($L=4$), 274 ($L=6$), 237 ($L=8$), and 202 ($L=12$), with corresponding $\hat{R}$ values of 1.004, 1.000, 1.000, 1.015, 1.020, 1.001, and 1.039.

\begin{figure}[ht!]
    \centering
    \includegraphics[width = \textwidth]{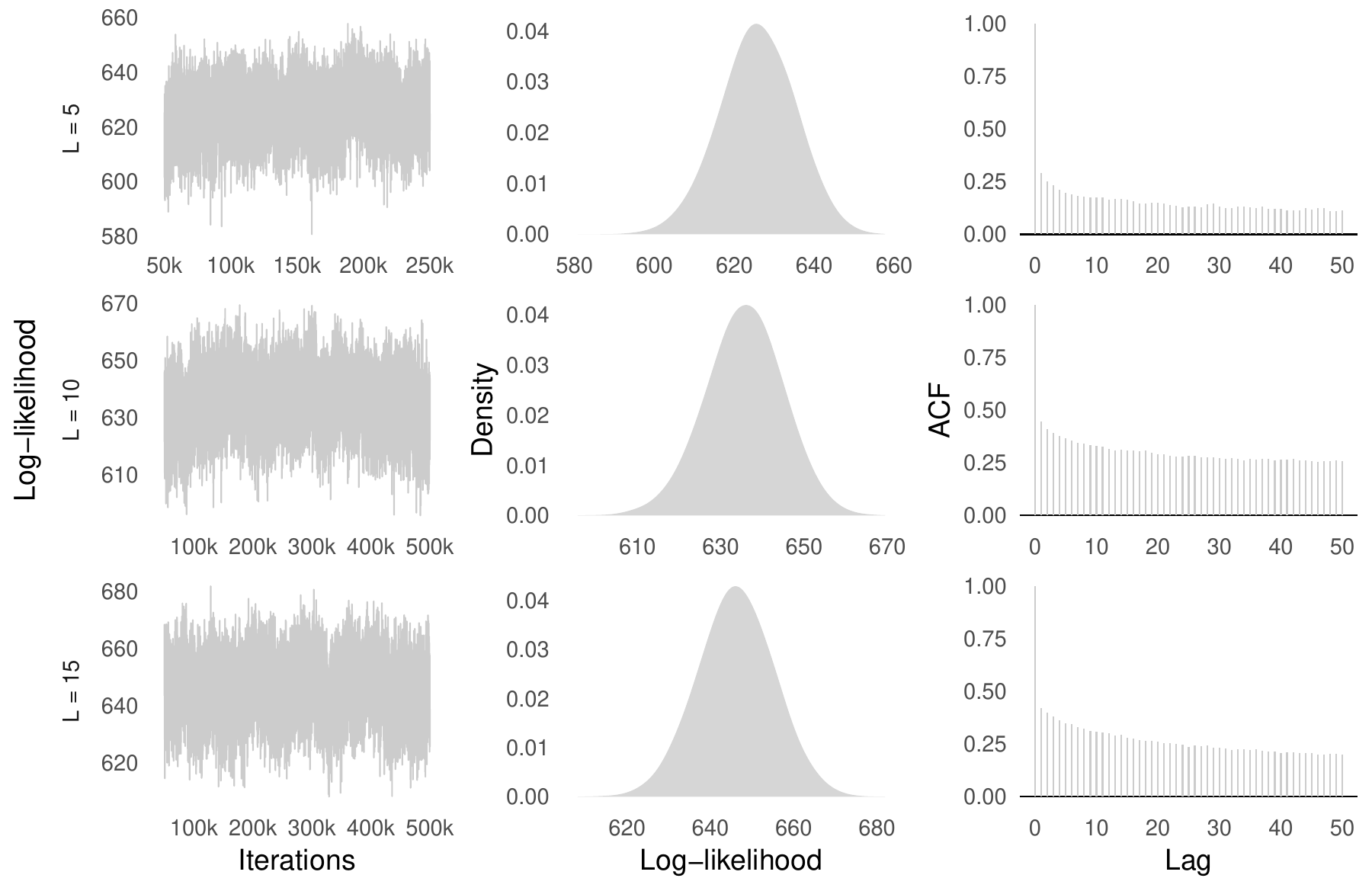}
    \caption{Trace plot, density estimates and autocorrelation function of the pseudo log-likelihood samples obtained for Example 4.}
    \label{fig__L10__convergence}
\end{figure}

Figure \ref{fig__L10__convergence} displays convergence diagnostics for Example 4. All chains indicate satisfactory convergence, with ESS per 10,000 samples of 387, 114, and 176 and $\hat{R}$ values of 1.000, 1.062, and 1.000 for $L = 5$, $10$, and $15$, respectively.

\begin{figure}[ht!]
    \centering
    \includegraphics[width = \textwidth]{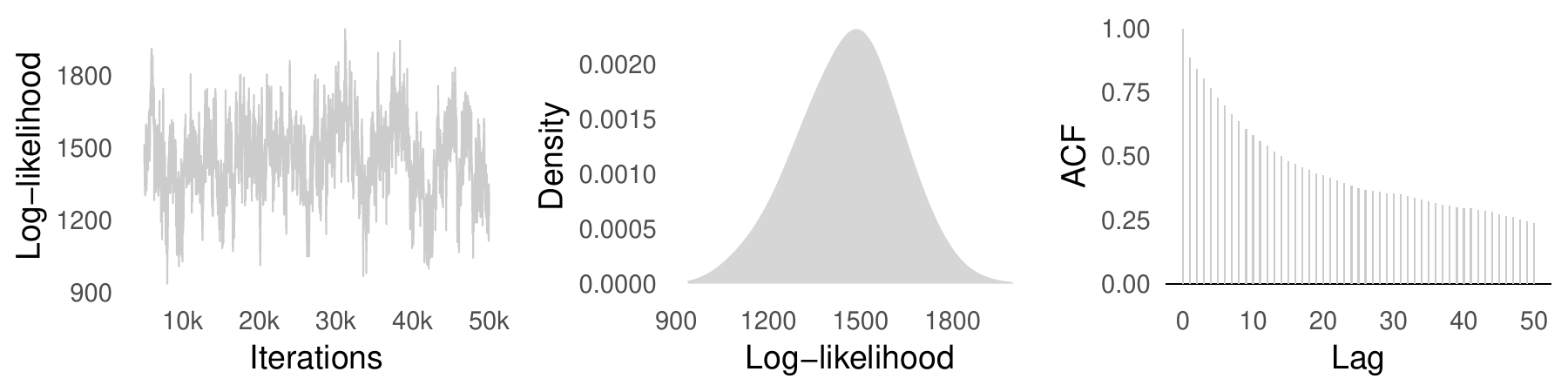}
    \caption{Trace plot, density estimates and autocorrelation function of the pseudo log-likelihood samples obtained for the Mato Grosso fires dataset.}
    \label{fig__fires__convergence}
\end{figure}

Figures \ref{fig__fires__convergence} present convergence diagnostics for the Mato Grosso fires dataset, with an ESS per 10,000 samples of 269 and $\hat{R}$ = 1.065.

\begin{thebibliography}{90}

\bibitem{adams2009tractable}
Adams, R. P., I. Murray \& D. J. MacKay (2009). Tractable nonparametric Bayesian inference in Poisson processes with Gaussian process intensities. In \textit{Proceedings of the 26th Annual International Conference on Machine Learning}, pp. 9--16.

\bibitem{albert1993bayesian}
Albert, J. H. \& S. Chib (1993). Bayesian analysis of binary and polychotomous response data. \textit{Journal of the American Statistical Association}, 88(422), 669--679.

\bibitem{aurenhammer2013voronoi}
Aurenhammer, F., R. Klein \& D.-T. Lee (2013). \textit{Voronoi Diagrams and Delaunay Triangulations}. World Scientific Publishing Company.

\bibitem{baddeley2000non}
Baddeley, A. J., J. M{\o}ller \& R. Waagepetersen (2000). Non- and semi-parametric estimation of interaction in inhomogeneous point patterns. \textit{Statistica Neerlandica}, 54(3), 329--350.

\bibitem{baddeley2015spatial}
Baddeley, A., E. Rubak \& R. Turner (2015). \textit{Spatial Point Patterns: Methodology and Applications with R}. Chapman \& Hall/CRC Press.


\bibitem{beskos2006exact}
Beskos, A., O. Papaspiliopoulos, G. O. Roberts \& P. Fearnhead (2006). Exact and computationally efficient likelihood-based estimation for discretely observed diffusion processes (with discussion). \textit{Journal of the Royal Statistical Society: Series B}, 68(3), 333--382.

\bibitem{brix2001spatiotemporal}
Brix, A. \& P. J. Diggle (2001). Spatiotemporal prediction for log-Gaussian Cox processes. \textit{Journal of the Royal Statistical Society: Series B}, 63(4), 823--841.

\bibitem{cotter2013mcmc}
Cotter, S. L., G. O. Roberts, A. M. Stuart \& D. White (2013). MCMC methods for functions: modifying old algorithms to make them faster. \textit{Statistical Science}, pages 424--446.

\bibitem{cox1955some}
Cox, D. R. (1955). Some statistical methods connected with series of events. \textit{Journal of the Royal Statistical Society: Series B}, 17(2), 129--164.


\bibitem{silva2024exact}
da Silva, D. M. \& D. Gamerman (2024). Exact Bayesian geostatistics under preferential sampling. \textit{Bayesian Analysis}, pages 1--29.

\bibitem{dangelo2022local}
D'Angelo, N., M. Siino, A. D'Alessandro \& G. Adelfio (2022). Local spatial log-Gaussian Cox processes for seismic data. \textit{AStA Advances in Statistical Analysis}, 106(3), 633--671.

\bibitem{datta2016hierarchical}
Datta, A., S. Banerjee, A. O. Finley \& A. E. Gelfand (2016). Hierarchical nearest-neighbor Gaussian process models for large geostatistical datasets. \textit{Journal of the American Statistical Association}, 111(514), 800--812.

\bibitem{diggle1985kernel}
Diggle, P. (1985). A kernel method for smoothing point process data. \textit{Journal of the Royal Statistical Society: Series C}, 34(2), 138--147.

\bibitem{diggle2014statistical}
Diggle, P. J. (2014). \textit{Statistical Analysis of Spatial and Spatio-Temporal Point Patterns}. CRC Press.

\bibitem{diggle2013spatial}
Diggle, P., P. Moraga, B. Rowlingson \& B. Taylor (2013). Spatial and spatio-temporal log-Gaussian Cox processes: extending the geostatistical paradigm. \textit{Statistical Science}, 28, 542--563.

\bibitem{dunlop2017hierarchical}
Dunlop, M. M., M. A. Iglesias \& A. M. Stuart (2017). Hierarchical Bayesian level set inversion. \textit{Statistics and Computing}, 27, 1555--1584.

\bibitem{dvorak2019quick}
Dvo\v{r}\'ak, J. \& J. M{\o}ller (2019). Quick inference for log Gaussian Cox processes with non-stationary underlying random fields. \textit{Spatial Statistics}, 33, 100381.

\bibitem{rcpp2011}
Eddelbuettel, D. \& R. Fran\c{c}ois (2011). Rcpp: Seamless R and C++ integration. \textit{Journal of Statistical Software}, 40(8), 1--18.

\bibitem{fruhwirth2011dealing}
Fr{\"u}hwirth-Schnatter, S. (2011). Dealing with label switching under model uncertainty. In \textit{Mixtures: Estimation and Applications}, pp. 213--239. Wiley Online Library.



\bibitem{gonccalves2023exactlevelset}
Gon\c{c}alves, F. B. \& B. C. Dias (2023). Exact Bayesian inference for level-set Cox processes with piecewise constant intensity function. \textit{Journal of Computational and Graphical Statistics}, 32(1), 1--18.

\bibitem{gonccalves2024likelihood}
Gon\c{c}alves, F. B. \& P. Franklin (2024). Likelihood function: definition, examples, and numerical experiments. \textit{Chilean Journal of Statistics}, 15(2).

\bibitem{gonccalves2018exact}
Gon\c{c}alves, F. B. \& D. Gamerman (2018). Exact Bayesian inference in spatiotemporal Cox processes driven by multivariate Gaussian processes. \textit{Journal of the Royal Statistical Society: Series B}, 80(1), 157--175.

\bibitem{gonccalves2023corrigendum}
Gon\c{c}alves, F. B. \& D. Gamerman (2023). Corrigendum: Exact Bayesian inference in spatiotemporal Cox processes driven by multivariate Gaussian processes. \textit{Journal of the Royal Statistical Society: Series B}, 85(1), 176--176.


\bibitem{gonccalves2023exactmontecarlo}
Gon\c{c}alves, F. B., K. Latuszyński \& G. O. Roberts (2023). Exact Monte Carlo likelihood-based inference for jump-diffusion processes. \textit{Journal of the Royal Statistical Society: Series B}, 85(3), 732--756.

\bibitem{gonccalves2022beyond}
Gon\c{c}alves, F. B., M. O. Prates \& G. A. S. Aguilar (2022). Beyond Gaussian processes: flexible Bayesian modeling and inference for geostatistical processes. \textit{arXiv preprint arXiv:2203.06437}.

\bibitem{goncalves2025pcnngp}
Gon\c{c}alves, F. B., M. O. Prates \& G. O. Roberts (2026). Bridging theory and practice in efficient Gaussian process-based statistical modeling for large datasets. \textit{arXiv preprint arXiv:2603.18324}.


\bibitem{heikkinen1998non}
Heikkinen, J. \& E. Arjas (1998). Non-parametric Bayesian estimation of a spatial Poisson intensity. \textit{Scandinavian Journal of Statistics}, 25(3), 435--450.

\bibitem{hildeman2018level}
Hildeman, A., D. Bolin, J. Wallin \& J. B. Illian (2018). Level set Cox processes. \textit{Spatial Statistics}, 28, 169--193.

\bibitem{hubbell1983diversity}
Hubbell, S. P. \& R. B. Foster (1983). Diversity of canopy trees in a neotropical forest and implications for conservation. In \textit{Tropical Rain Forest: Ecology and Management}, pp. 25--41. Oxford: Blackwell Scientific Publications.


\bibitem{illian2012toolbox}
Illian, J. B., S. H. S{\o}rbye \& H. Rue (2012). A toolbox for fitting complex spatial point process models using integrated nested Laplace approximation (INLA). \textit{Annals of Applied Statistics}, 6(4), 1499--1530.

\bibitem{bdqueimadas2025}
Instituto Nacional de Pesquisas Espaciais (INPE) (2025). BDQueimadas: Monitoramento de Queimadas e Incêndios (In Portuguese). Retrieved from \url{https://terrabrasilis.dpi.inpe.br/queimadas/bdqueimadas/}. Accessed in 2025-07-09.

\bibitem{kim2005analyzing}
Kim, H.-M., B. K. Mallick \& C. C. Holmes (2005). Analyzing nonstationary spatial data using piecewise Gaussian processes. \textit{Journal of the American Statistical Association}, 100(470), 653--668.

\bibitem{kingman1992poisson}
Kingman, J. F. C. (1992). \textit{Poisson Processes}, volume 3. Clarendon Press.


\bibitem{lee2023unified}
Lee, J., X. Miscouridou \& F. Caron (2023). A unified construction for series representations and finite approximations of completely random measures. \textit{Bernoulli}, 29(3), 2142--2166.

\bibitem{lewis1979simulation}
Lewis, P. W. \& G. S. Shedler (1979). Simulation of nonhomogeneous Poisson processes by thinning. \textit{Naval Research Logistics Quarterly}, 26(3), 403--413.

\bibitem{liang2009analysis}
Liang, S., B. P. Carlin \& A. E. Gelfand (2009). Analysis of Minnesota colon and rectum cancer point patterns with spatial and nonspatial covariate information. \textit{The Annals of Applied Statistics}, pages 943--962.

\bibitem{Lu06082025}
Lu, C., Y. Guan, M.-C. van Lieshout \& G. Xu (2025). XGBoostPP: Tree-based estimation of point process intensity functions. \textit{Journal of Computational and Graphical Statistics}, 0(0), 1--13.

\bibitem{luo2024nonstationary}
Luo, Z. T., H. Sang \& B. Mallick (2024). A nonstationary soft partitioned Gaussian process model via random spanning trees. \textit{Journal of the American Statistical Association}, 119(547), 2105--2116.

\bibitem{moller1998log}
M{\o}ller, J., A. R. Syversveen \& R. P. Waagepetersen (1998). Log Gaussian Cox processes. \textit{Scandinavian Journal of Statistics}, 25(3), 451--482.

\bibitem{moller2004statistical}
M{\o}ller, J. \& R. P. Waagepetersen (2004). \textit{Statistical Inference and Simulation for Spatial Point Processes}. Chapman \& Hall/CRC Press.

\bibitem{moller2007modern}
M{\o}ller, J. \& R. P. Waagepetersen (2007). Modern statistics for spatial point processes. \textit{Scandinavian Journal of Statistics}, 34(4), 643--684.

\bibitem{nguyen2024independent}
Nguyen, T. D., J. Huggins, L. Masoero, L. Mackey \& T. Broderick (2024). Independent finite approximations for Bayesian nonparametric inference. \textit{Bayesian Analysis}, 19(4), 1187--1224.

\bibitem{paciorek2006spatial}
Paciorek, C. J. \& M. J. Schervish (2006). Spatial modelling using a new class of nonstationary covariance functions. \textit{Environmetrics}, 17(5), 483--506.


\bibitem{papaspiliopoulos2008retrospective}
Papaspiliopoulos, O. \& G. O. Roberts (2008). Retrospective Markov chain Monte Carlo methods for Dirichlet process hierarchical models. \textit{Biometrika}, 95(1), 169--186.

\bibitem{geobr2024}
Pereira, R. H. M. \& C. N. Gon\c{c}alves (2024). \textit{geobr: Download Official Spatial Data Sets of Brazil}. R Foundation for Statistical Computing. R package version 1.9.1.


\bibitem{pinto2015point}
Pinto Junior, J. A., D. Gamerman, M. S. Paez \& R. H. Fonseca Alves (2015). Point pattern analysis with spatially varying covariate effects, applied to the study of cerebrovascular deaths. \textit{Statistics in Medicine}, 34(7), 1214--1226.

\bibitem{polson2013}
Polson, N. G., J. G. Scott \& J. Windle (2013). Bayesian inference for logistic models using P\'olya--Gamma latent variables. \textit{Journal of the American Statistical Association}, 108(504), 1339--1349.

\bibitem{pope2021gaussian}
Pope, C. A., J. P. Gosling, S. Barber, J. S. Johnson, T. Yamaguchi, G. Feingold \& P. G. Blackwell (2021). Gaussian process modeling of heterogeneity and discontinuities using Voronoi tessellations. \textit{Technometrics}, 63(1), 53--63.

\bibitem{quinlan2021class}
Quinlan, J. J., F. A. Quintana \& G. L. Page (2021). On a class of repulsive mixture models. \textit{Test}, 30, 445--461.

\bibitem{R2025}
R Core Team (2025). \textit{R: A Language and Environment for Statistical Computing}. Vienna, Austria: R Foundation for Statistical Computing.

\bibitem{ripley1977modelling}
Ripley, B. D. (1977). Modelling spatial patterns. \textit{Journal of the Royal Statistical Society: Series B}, 39(2), 172--192.

\bibitem{sampson1992nonparametric}
Sampson, P. D. \& P. Guttorp (1992). Nonparametric estimation of nonstationary spatial covariance structure. \textit{Journal of the American Statistical Association}, 87(417), 108--119.


\bibitem{lgcppackage}
Taylor, B. M., T. M. Davies, B. S. Rowlingson \& P. J. Diggle (2013). lgcp: An R package for inference with spatial and spatio-temporal log-Gaussian Cox processes. \textit{Journal of Statistical Software}, 52(4), 1--40.

\bibitem{tierney1998note}
Tierney, L. (1998). A note on Metropolis-Hastings kernels for general state spaces. \textit{Annals of Applied Probability}, 8, 1--9.

\bibitem{webster1992sample}
Webster, R. \& M. A. Oliver (1992). Sample adequately to estimate variograms of soil properties. \textit{Journal of Soil Science}, 43(1), 177--192.

\end{thebibliography}
\end{document}